\documentclass[aps,prb,amssymb,showpacs,twocolumn,tightenlines]{revtex4}
\usepackage{graphicx}
\usepackage{amsmath,bm}
\usepackage[mathscr]{euscript}
\usepackage{dcolumn}

\begin{document}
\title{Emergence of an excitonic collective mode in the dilute electron gas}


\author{Yasutami Takada}
\thanks{Email: takada@issp.u-tokyo.ac.jp: This paper will be published 
in Phys. Rev. B.}

\affiliation{Institute for Solid State Physics, University of Tokyo, 
Kashiwa, Chiba 277-8581, Japan}

\date{\today}    
\begin{abstract}
By comparing two expressions for the polarization function $\Pi({\bm q},i\omega)$ 
given in terms of two different local-field factors, $G_+({\bm q},i\omega)$ and 
$G_s({\bm q},i\omega)$, we have derived the kinetic-energy-fluctuation (or 
sixth-power) sum rule for the momentum distribution function $n({\bm p})$ in the 
three-dimensional electron gas. With use of this sum rule, together with the 
total-number (or second-power) and the kinetic-energy (or fourth-power) sum rules, 
we have obtained $n({\bm p})$ in the low-density electron gas at negative 
compressibility (namely, $r_s>5.25$ with $r_s$ being the conventional density 
parameter) up to $r_s \approx 22$ by improving on the interpolation scheme due to 
Gori-Giorge and Ziesche proposed in 2002. The obtained results for $n({\bm p})$ 
combined with the improved form for $G_s({\bm q},\omega\!+\!i0^+)$ are employed to 
calculate the dynamical structure factor $S({\bm q},\omega)$ to reveal that a giant 
peak, even bigger than the plasmon peak, originating from an excitonic collective 
mode made of electron-hole pair excitations, emerges in the low-$\omega$ region at 
$|{\bm q}|$ near $2p_{\rm F}$ ($p_{\rm F}$: the Fermi wave number). Connected 
with this mode, we have discovered a singular point in the retarded dielectric 
function at $\omega\!=\!0$ and $|{\bm q}|\!\approx\! 2p_{\rm F}$. 
\end{abstract}


\maketitle

\section{Introduction}
\label{sec:1}

The electron gas, an assembly of $N$ electrons embedded in a uniform positive rigid 
background, has been investigated for a very long time, not only because it is 
relevant in understanding the exchange-correlation effects in simple metals 
and semiconductors, but also because it provides indispensable information for the 
actual implementation of the density functional theory (DFT), especially in its 
local density approximation (LDA).~\cite{GV2005}

From a physical point of view, we may claim that this system is even more important 
in the low-density regime (namely, $p_{\rm F}a_0 \ll 1$ with $p_{\rm F}$ the Fermi 
wave number and $a_0$ the lattice constant), because in this regime the 
first-principles Hamiltonian to describe conduction electrons in a metal is 
universally reduced to the one for the electron gas, leading us to recognize that 
understanding the physical properties of the dilute electron gas is nothing but 
revealing the {\it universal} electron-correlation behavior of actual dilute metals.

The dilute electron gas, however, is not well understood; the thermodynamic 
quantities, such as the correlation energy $\varepsilon_c$ and the compressibility 
$\kappa$, are accurately obtained as a function of the density parameter $r_s$ by 
the Green's-Function Monte Carlo (GFMC) method~\cite{CA1980} and the interpolation 
formulas to reproduce the GFMC data~\cite{VWN1980,PW1992}, where $r_s$ is defined by 
$r_s\!=\!(4\pi n/3)^{-1/3}/a_B\!=\!(\alpha p_{\rm F} a_B)^{-1}$ with $n\!=\!N/\Omega_t$ 
the electron density, $\Omega_t$ the total volume of the system, $\alpha\!=\!
(4/9\pi)^{1/3}\!\approx \! 0.5211$ and $a_B$ the Bohr radius. The diffusion Monte 
Carlo (DMC) calculations are done to analyze the ground-state phases, including spin 
polarization, in  the wide range of $r_s$~\cite{Ortiz1999} and recently for 
$0.5 \!\leq r_s \!\leq 20$ in more detail~\cite{Needs2013}, but we do not know the 
precise behavior of other physical quantities, such as the momentum distribution 
function $n({\bm p})$ for $r_s\!>\!5$. It is true that some quantum Monte Carlo 
calculations are done to obtain $n({\bm p})$ for $r_s \!\leq \!
10$,~\cite{Ortiz1994,Holtzmann2011} but the results are not very accurate 
due probably to large size effects and improper starting trial functions, 
judging from the assessment of their accuracy by the sum rules for 
$n({\bm p})$.~\cite{YT1991a,Maebashi2011} As for the dynamical structure factor 
$S({\bm q},\omega)$, no reliable data are available in the dilute metallic regime.

One of the characteristic features of the dilute electron gas is the appearance of 
``dielectric catastrophe'' associated with the divergence of $\kappa$ at 
$r_s=r_s^c\!\equiv\! 5.25$ followed by negative $\kappa$ for 
$r_s>r_s^c$,~\cite{Mahan2000,YT1991} implying that the static polarization function 
$\Pi({\bm q},0)$ in the long wave-length limit is also {\it negative} (and thus 
dielectrically anomalous) owing to the compressibility sum rule, $\lim_{|{\bm q}|
\to 0}\Pi({\bm q},0)\!=\!n^2 \kappa \Omega_t$.~\cite{Pines1966} This anomaly for 
$r_s>r_s^c$ brings about the curious behavior of the ion-ion pair correlation 
function in expanded liquid alkali 
metals,~\cite{Matsuda2007,Maebashi2009,Maebashi2009a} but it does not induce any 
instabilities in the electron gas itself, as long as $1/\kappa$ changes continuously 
from positive to negative through the point of $1/\kappa\!=\!0$,~\cite{Dolgov1981} 
because the electron-electron effective interaction is not determined only by the 
dielectric function $\varepsilon({\bm q},i\omega)\!=\!1\!+\!V({\bm q})
\Pi({\bm q},i\omega)$ with $V({\bm q})\!=\!4\pi e^2/(\Omega_t {\bm q}^2)$ the bare 
Coulomb interaction,~\cite{Kukkonen1979} excluding the occurrence of CDW-type 
instabilities at $r_s=r_s^c$.~\cite{Schakel2001} 

By analyzing $\varepsilon^R({\bm q},\omega) [\!=\!\varepsilon({\bm q},\omega\!+
\!i0^+)]$ the retarded dielectric function as $r_s$ approaches $r_s^c$ from the 
positive side of $\kappa$ (or $r_s\!<\!r_s^c$), we identify the physical origin of 
this divergence of $\kappa$ as the ``enhanced excitonic effect'' or the effect of 
strong electron-hole attraction in an electron-hole single-pair 
excitation.~\cite{YT2005} For $r_s\!>\!r_s^c$, by considering causality, Takayanagi 
and Lipparini~\cite{Takayanagi1997} conclude that $\Pi({\bm q},i\omega)$ contains an 
anomalous part $\Pi_a({\bm q},i\omega)$ made of a single pair of poles at $i\omega=
\pm i\tilde{\omega}_{\rm ex}({\bm q})$ in the form of 
\begin{align}
\label{eq:1}
\Pi_a({\bm q},i\omega)
=\frac{1}{V({\bm q})}\frac{ne^2}{m}
\frac{f_{\rm ex}({\bm q})}{\omega^2-\tilde{\omega}_{\rm ex}({\bm q})^2},
\end{align}
with $m$ the mass of a free electron. These poles at which $\Pi({\bm q},i\omega)$ 
diverges, however, do not give any prominent structures in the observable physical 
quantity $S({\bm q},\omega)$, which can be calculated by $S({\bm q},\omega)=
-{\rm Im}Q_c^R({\bm q},\omega)/\pi$ for $\omega>0$ at zero temperature ($T=0$) 
with the retarded charge response function $Q_c^R({\bm q},\omega)$, given by
\begin{align}
\label{eq:2}
Q_c^R({\bm q},\omega)=-\frac{\Pi({\bm q},\omega\!+\!i0^+)}
{1+V({\bm q})\Pi({\bm q},\omega\!+\!i0^+)}.
\end{align}
Then a couple of questions arise: what is the physical role of these poles and 
how the strong excitonic effect manifests itself in $S({\bm q},\omega)$ for 
$r_s>r_s^c$? We shall address these questions by accurately determining 
$\tilde{\omega}_{\rm ex}({\bm q})$, ``the oscillator strength'' of the pole 
$f_{\rm ex}({\bm q})$, and $S({\bm q},\omega)$ for $r_s$ from $r_s^c$ up to about 20 
in which the ground state has already been confirmed to be a paramagnetic 
metal.~\cite{Needs2013}

A straightforward way of obtaining accurate results for $S({\bm q},\omega)$  is to 
perform the highly self-consistent calculation in the GW$\Gamma$ scheme~\cite{YT2001} 
which includes the vertex function $\Gamma$ satisfying the Ward identity 
(WI),~\cite{Ward1950,Takahashi1957} just as done for $S({\bm q},\omega)$ for 
$r_s\leq 5$,~\cite{YT2002} but it turns out that this GW$\Gamma$ does not work 
well in the dielectric-catastrophe regime. Thus it is revised into the 
G$\tilde{{\rm W}}\Gamma_{\rm WI}$ scheme~\cite{Maebashi2011,YT2016} to obtain the 
self-consistent results for $S({\bm q},\omega)$ as well as $n({\bm p})$ up to $r_s
=10$, but there still remains the problem of reaching a fully self-consistent 
solution for $r_s>10$. 

Confronted with this difficulty, we will take the following strategy; firstly, we 
reconsider the parametrization scheme for determining $n({\bm p})$ due to Gori-Giorgi 
and Ziesche (GZ),~\cite{Ziesche2002} who interpolate the accurate data for 
$n({\bm p})$ in the effective potential expansion (EPX) method~\cite{YT1991a} for 
$1 \!\leq \!r_s \! \leq 5$ with that in the Wigner-crystal limit ($r_s\! \gg \!10$), 
together with the two sum rules, one for the total number and the other for the 
total kinetic energy,~\cite{YT1991a} expressed as
\begin{align}
\label{eq:3}
\sum_{{\bm p}\sigma}n({\bm p})&=N,
\\
\label{eq:4}
\sum_{{\bm p}\sigma}\varepsilon_{\bm p}n({\bm p})
&=N\langle \varepsilon_{\bm p} \rangle 
=N\left (\frac{3}{5}E_{\rm F}-\varepsilon_c-
r_s\frac{\partial \varepsilon_c}{\partial r_s}\right ),
\end{align}
with $\sigma$ the spin variable, $\varepsilon_{\bm p}\!=\!{\bm p}^2/2m$, and 
$E_{\rm F}\!=\!p_{\rm F}^2/2m$ the Fermi energy. This GZ scheme can give rather 
accurate $n({\bm p})$ for $r_s\!\lesssim \!12$. Here we shall 
improve on it by increasing the number of input data for $n({\bm p})$ that we obtain 
up to $r_s=10$ in G$\tilde{{\rm W}}\Gamma_{\rm WI}$. We also use, in addition to the 
above two, the third sum rule for the total kinetic-energy fluctuation, written as
\begin{align}
\label{eq:5}
\sum_{{\bm p}\sigma}\varepsilon_{\bm p}^2n({\bm p})
&=N \left [ \langle (\varepsilon_{\bm p}-\langle \varepsilon_{\bm p} \rangle )^2 
\rangle +\langle \varepsilon_{\bm p}\rangle ^2 \right ].
\end{align}
We shall derive this sum rule in Sec.~\ref{sec:2} with giving a concrete value 
for the righthand side of Eq.~(\ref{eq:5}).

Secondly, once $n({\bm p})$ is known, we can calculate $\Pi({\bm q},i\omega)$ by 
the formula given in G$\tilde{{\rm W}}\Gamma_{\rm WI}$ as
\begin{align}
\label{eq:6}
\Pi({\bm q},i\omega)=\frac{\Pi_{\rm WI}({\bm q},i\omega)}
{1-V({\bm q})G_s({\bm q},i\omega)\Pi_{\rm WI}({\bm q},i\omega)},
\end{align}
where $\Pi_{\rm WI}({\bm q},i\omega)$ is defined as
\begin{align}
\label{eq:7}
\Pi_{\rm WI}({\bm q},i\omega)=\sum_{{\bm p}\sigma}
\frac{n({\bm p}+{\bm q})-n({\bm p})}
{i\omega-\varepsilon_{{\bm p}+{\bm q}}+\varepsilon_{\bm p}},
\end{align}
with $G_s({\bm q},i\omega)$ the local-field factor as introduced by Richardson and 
Ashcroft.~\cite{Richardson1994} This factor is different from the conventinal 
local-field factor $G_+({\bm q},i\omega)$,~\cite{Kukkonen1979} with which 
$\Pi({\bm q},i\omega)$ is expressed in another way as
\begin{align}
\label{eq:8}
\Pi({\bm q},i\omega)=\frac{\Pi_{0}({\bm q},i\omega)}
{1-V({\bm q})G_+({\bm q},i\omega)\Pi_{0}({\bm q},i\omega)},
\end{align}
with $\Pi_{0}({\bm q},i\omega)$ the Lindhard function, given by
\begin{align}
\label{eq:9}
\Pi_{0}({\bm q},i\omega)=\sum_{{\bm p}\sigma}
\frac{n_0({\bm p}+{\bm q})-n_0({\bm p})}
{i\omega-\varepsilon_{{\bm p}+{\bm q}}+\varepsilon_{\bm p}},
\end{align}
with $n_0({\bm p})=\theta(p_{\rm F}-|{\bm p}|)$ the step function. The obtained 
$\Pi({\bm q},i\omega)$ can be used to determine both $\tilde{\omega}_{\rm ex}({\bm q})$ 
and $f_{\rm ex}({\bm q})$ by the analysis of its divergent property. 

Finally, we need to make an analytic continuation of $\Pi({\bm q},i\omega)$ into 
$\Pi({\bm q},\omega\!+\!i0^+)$ to calculate $S({\bm q},\omega)$. There are two 
ingredients, $\Pi_{\rm WI}({\bm q},\omega\!+\!i0^+)$ and $G_s({\bm q},\omega\!+\!i0^+)$, 
in this analytic continuation. The former can be obtained by the direct calculation 
through Eq.~(\ref{eq:7}) with $\omega\!+\!i0^+$ in place of $i\omega$. As for the 
latter, we find it better to slightly modify $G_s({\bm q},i\omega)$ from the 
original form~\cite{Richardson1994}, so that we can not only obtain much more 
accurate $\varepsilon_c$ in a wide range of $r_s$ by the adiabatic connection 
formula~\cite{Lein2000} but also eliminate unphysical divergences from 
$G_s({\bm q},\omega\!+\!i0^+)$.

\begin{figure}[htbp]
\begin{center}
\includegraphics[scale=0.525,keepaspectratio]{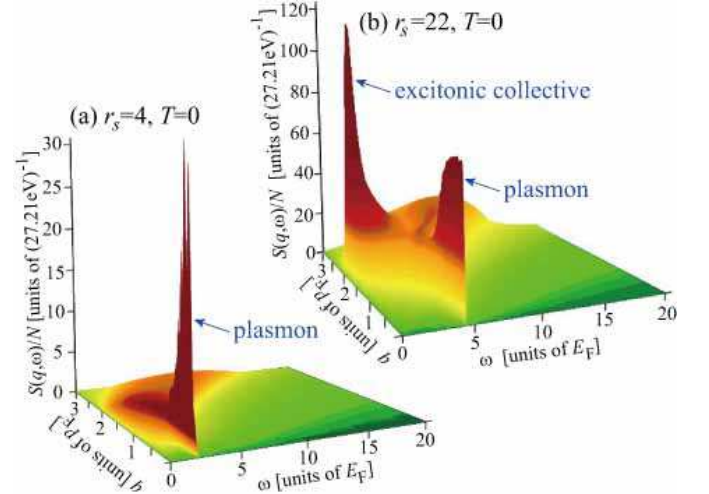}
\end{center}
\caption[Fig.1]{(Color online) Bird's-eye view of $S({\bm q},\omega)$ in the electron 
gas at zero temperature at (a) a usual metallic density $r_s=4$, in which the 
plasmon peak dominates, and (b) a very dilute density $r_s=22$, in which 
an excitonic collective peak, much bigger than the plasmon peak, emerges.}
\label{fig:1}
\end{figure}

In accordance with the above strategy, we have successfully performed the calculation 
and obtained the results of both $n({\bm p})$ and $S({\bm q},\omega)$ for $r_s$ 
up to over 20 to reveal that a giant peak other than the conventional plasmon one 
emerges in $S({\bm q},\omega)$ for $r_s \agt 10$, evolving into a predominant peak 
for $r_s>20$, as shown in Fig.~\ref{fig:1}. Concomitantly a singular behavior 
is found in $n({\bm p})$ for $r_s\agt 22$, which might suggest the occurrence 
of an electronic phase transition, though at present we cannot draw a definite 
conclusion on this point. Because the new peak in $S({\bm q},\omega)$ is evolved 
from the shoulder structure due to the excitonic effect found at usual 
metallic densities~\cite{YT2002}, this peak may be called ``an excitonic collective 
mode'', a new concept in the dilute electron gas. In the dielectric-catastrophe 
regime, we also find that $\varepsilon^R({\bm q},\omega)$ is controlled by the value 
of ${\bm q}_{\rm ex}(\neq {\bm 0})$ at which $\tilde{\omega}_{\rm ex}({\bm q}_{\rm ex})=0$, 
bringing about a singular point in $\varepsilon^R({\bm q},\omega)$ at $\omega=0$ and 
${\bm q}={\bm q}_{\rm ex}$ for $r_s>r_s^c$, a feature never seen in the conventional 
metals like the electron gas for $r_s<r_s^c$. 

This paper is organized as follows: In Sec.~\ref{sec:2} we derive the third sum 
rule for $n({\bm p})$, with which we improve on the GZ scheme to calculate 
$n({\bm p})$ in the electron gas. Our improved scheme works very well to obtain 
$n({\bm p})$ up to $r_s\approx 22$. In Sec.~\ref{sec:3} we 
calculate $S({\bm q},\omega)$ with use of the obtained $n({\bm p})$ and the local-field 
factor $G_s({\bm q},\omega\!+\!i0^+)$, the latter of which is improved on the original 
form of Richardson and Ashcroft (RA). For $r_s$ larger than about 10, a novel peak 
structure appears in $S({\bm q},\omega)$ and it dominates the plasmon peak for $r_s>20$. 
The origin of this peak is investigated by a detailed study of 
$\varepsilon^R({\bm q},\omega)$. The results obtained in this paper are summarized 
in Sec.~\ref{sec:4}, together with discussions on related issues. In this paper, 
we employ units in which $\hbar=k_{\rm B}=1$.

\section{Momentum Distribution Function}
\label{sec:2}

\subsection{Hamiltonian}
\label{sec:2A}

Apart from a constant term, the first-principles Hamiltonian $H$ for a system of $N$ 
electrons interacting with each other through the Coulomb interaction in a periodic 
one-body potential is written in the plane-wave basis as~\cite{YT1993}
\begin{align}
\label{eq:10}
H=H_0+V+U,
\end{align}
with
\begin{align}
\label{eq:11}
H_0&=\sum_{{\bm p}\sigma}\varepsilon_{{\bm p}}c_{{\bm p}\sigma}^+ c_{{\bm p}\sigma},
\\
\label{eq:12}
V&=\frac{1}{2}\sum_{{\bm q}\neq {\bm 0}}\sum_{{\bm p}\sigma}
\sum_{{\bm p'}\sigma'}V({\bm q})c_{{\bm p}+{\bm q}\sigma}^+
c_{{\bm p'}-{\bm q}\sigma'}^+ c_{{\bm p'}\sigma'}c_{{\bm p}\sigma},
\\  
\label{eq:13}
U&=\sum_{{\bm p}\sigma}\sum_{{\bm G}\neq {\bm 0}}U_{{\bm p},{\bm p}+{\bm G}}
c_{{\bm p}\sigma}^+ c_{{\bm p}+{\bm G}\sigma},
\end{align}
where $c_{{\bm p}\sigma}$ is the annihilation operator of an electron with 
momentum ${\bm p}$ and spin $\sigma$ and ${\bm G}$ is a reciprocal-lattice vector. 
If the electron density $n (=N/\Omega_t)$ is so low as to satisfy the condition of 
$p_{\rm F}a_0 \ll 1$, we may virtually regard $|{\bm G}|$, which is $2\pi/a_0$ 
or larger, as an infinite number, allowing us to neglect all terms in $U$. Then the 
system can be decribed by the Hamiltonian reduced in the form of $H=H_0+V$, which 
is nothing but the electron gas and with which we shall be concerned in this paper. 

\subsection{Momentum distribution function}
\label{sec:2B}

At zero temperature ($T\!=\!0$), the momentum distribution function $n({\bm p})$ is 
defined as an expectation value 
\begin{align}
n({\bm p})=\langle c_{{\bm p}\sigma}^+ c_{{\bm p}\sigma} \rangle, 
\label{eq:14}
\end{align}
evaluated with respect to the ground state $\Psi_0$. This quantity is independent of 
spin for a paramagnetic $\Psi_0$. 

\begin{figure}[htbp]
\begin{center}
\includegraphics[scale=0.38,keepaspectratio]{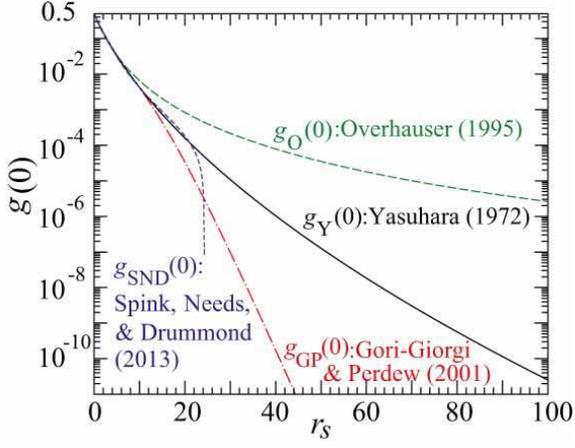}
\end{center}
\caption[Fig.2]{(Color online) Comparison of various formulas for the on-top pair 
distribution function $g(0)$ plotted as a function of $r_s$ over a very wide range.}
\label{fig:2}
\end{figure}

The short-range electron correlation determines the asymptotic behavior of 
$n({\bm p})$ as well as the structure factor $S({\bm q})$ and their exact forms are 
given as~\cite{Kimball1975}
\begin{align}
\label{eq:15}
&\Omega_t \left (\frac{a_B}{4\pi}\right )^2 \lim_{|\bm p|\to \infty}
\Bigl [|\bm p|^8 n({\bm p})\Bigr ]
\nonumber \\
&=\left (\frac{a_B}{8\pi}\right )
\lim_{|\bm q|\to \infty}\Bigl [|\bm q|^4 S({\bm q})\Bigr ]=g(0),
\end{align}
where $g(0)$ is the on-top value of the pair distribution function. We can rewite 
Eq.~(\ref{eq:15}) as
\begin{align}
\label{eq:16}
\lim_{|\bm p|\to \infty}n({\bm p})
=\frac{8}{9}\left ( \frac{\alpha r_s}{\pi}\right )^2
\frac{p_{\rm F}^8}{|\bm p|^8}\,g(0),
\end{align}
dictating the asymptotic behavior of $n({\bm p})$. By using $V({\bm q})$ in the 
electron-electron ladder approximation, Yasuhara derived a formula for $g(0)$ 
as~\cite{Yasuhara1972,Yasuhara1974}
\begin{align}
\label{eq:17}
g(0)=g_{\rm Y}(0) \equiv \frac{1}{2}
\left (\frac{\sqrt{4\alpha r_s/\pi}}{I_1(2\sqrt{4\alpha r_s/\pi})}\right )^2,
\end{align}
where $I_1(x)$ is the modified Bessel function of the first kind. There are three 
other parametrization formulas for $g(0)$; by the phase-shift analysis of the 
two-electron problem, Overhauser~\cite{Overhauser1995} devived one formula 
$g_{\rm O}(0)$ and Gori-Giorgi and Perdew~\cite{GP2001} improved on it to give 
another one $g_{\rm GP}(0)$, both of which are intended to provide accurate results 
of $g(0)$ for $r_s \leq 10$. More recently, Spink, Needs, and Drummond 
(SND)~\cite{Needs2013} provided yet another formula $g_{\rm SND}(0)$, which 
reproduces their DMC data for $r_s \leq 20$. The results of those formulas are given 
as a function of $r_s$ in Fig.~\ref{fig:2}. We see immediately that they are 
essentially the same for $r_s \leq 10$. For $r_s \leq 20$, Yasuhara's and SND's give 
about the same results, but the latter deteriorates very much soon after $r_s$ goes 
beyond $20$. (Actually, SND gives unphysical negative $g(0)$ for $r_s>24.26$.) Thus, 
apparently, the best choice for $g(0)$ is the Yasuhara's formula $g_{\rm Y}(0)$ in 
Eq.~(\ref{eq:17}) which will be mainly employed over a wide range of $r_s$ hereafter. 

\subsection{Sum rules}
\label{sec:2C}

Let us derive the three sum rules, Eqs.~(\ref{eq:3})-(\ref{eq:5}), for $n({\bm p})$. 
The static charge response function $Q_c^R({\bm q},0)$ in the electron gas is calculated 
by Moroni, Ceperley, and Senatore (MCS)~\cite{Moroni1995} in DMC for $r_s \leq 10$, 
giving the numerical data for $\Pi({\bm q},0)$. With use of Eq.~(\ref{eq:8}), we obtain 
\begin{align}
\label{eq:18}
\frac{1}{\Pi({\bm q},0)}=\frac{1}{\Pi_0({\bm q},0)}-V({\bm q})G_+({\bm q},0).
\end{align}
This may be regarded as the definition of the static local-field factor 
$G_+({\bm q},0)$. The DMC data of $\Pi({\bm q},0)$, together with the well-known 
analytical result of $\Pi_0({\bm q},0)$, determine $G_+({\bm q},0)$, which 
is parametrized as a function of $r_s$ and $q \,(\!=\!|{\bm q}|)$.~\cite{Moroni1995} 
In particular, in the limit of $q\!\to \! \infty$, $G_+({\bm q},0)$ is known as 
\begin{align}
\label{eq:19}
G_+({\bm q},0) \xrightarrow[q \to \infty]{}
C(r_s)\left (\frac{q}{p_{\rm F}} \right )^2+B(r_s),
\end{align}
where the constants $B(r_s)$ and $C(r_s)$ are given by
\begin{align}
\label{eq:20}
& B(r_s)=B_{\rm MCS}(r_s)\equiv \frac{1+(a_1+a_2r_s)\sqrt{r_s}}
{3+(b_1+b_2 r_s)\sqrt{r_s}},
\\
\label{eq:21}
&C(r_s)=\frac{\pi}{4}\alpha r_s \left ( -\varepsilon_c(r_s)
-r_s\frac{\partial \varepsilon_c(r_s)}{\partial r_s}\right ),
\end{align}
with $a_1=2.15$, $a_2=0.435$, $b_1=1.57$, $b_2=0.409$, and $\varepsilon_c(r_s)$ 
(in Ry) in the parameterization scheme due to Perdew and Wang.~\cite{PW1992} 

On the other hand, we obtain another expression for $\Pi({\bm q},0)^{-1}$ with use of 
Eq.~(\ref{eq:6}) as
\begin{align}
\label{eq:22}
\frac{1}{\Pi({\bm q},0)}=\frac{1}{\Pi_{\rm WI}({\bm q},0)}-V({\bm q})G_s({\bm q},0), 
\end{align}
where the other local-field factor $G_s({\bm q},0)$ was investigated by 
Niklasson~\cite{Niklasson1974} through the equation of motion method and is known 
to behave asymptotically as
\begin{align}
\label{eq:23}
G_s({\bm q},0) \xrightarrow[q \to \infty]{}
\frac{2}{3}\bigl [1-g(0) \Bigr ].
\end{align}
By combining Eq.~(\ref{eq:22}) with Eq.~(\ref{eq:18}), we obtain
\begin{align}
\label{eq:24}
\frac{1}{\Pi_{\rm WI}({\bm q},0)}
\!=\!\frac{1}{\Pi_0({\bm q},0)}
\!+\!\left [G_s({\bm q},0)\! -\!G_+({\bm q},0)\right ]V({\bm q}).
\end{align}

Now $\Pi_{\rm WI}({\bm q},0)$ in Eq.~(\ref{eq:7}) is reduced to
\begin{align}
\label{eq:25}
\Pi_{\rm WI}({\bm q},0)=\Omega_t\frac{mp_{\rm F}}{2\pi^2}
\int_0^{\infty}dx\,n(x)\,\frac{x}{z}\,\ln\left |\frac{x+z}{x-z}\right |,
\end{align}
with $x=|{\bm p}|/p_{\rm F}$, $z=|{\bm q}|/2p_{\rm F}$, and $n(x)=n({\bm p})$. 
This equation can be asymptotically expanded as
\begin{align}
\label{eq:26}
\Pi_{\rm WI}({\bm q},0)=\Omega_t\frac{mp_{\rm F}}{\pi^2}
\left (\frac{I_2}{z^2}+\frac{1}{3}\frac{I_4}{z^4}+
\frac{1}{5}\frac{I_6}{z^6}+\cdots \right ),
\end{align}
where the $n$th-power integral $I_n$ is defined as
\begin{align}
\label{eq:27}
I_n=\int_0^{\infty}dx\,n(x)\,x^n.
\end{align}
A similar asymptotic expansion can be made for $\Pi_{0}({\bm q},0)$ with $I_n$ 
replaced by $I_n^{(0)}=1/(n+1)$. Then, by comparing the both sides of 
Eq.~(\ref{eq:24}) order by order and taking the large-$q$ limit, we obtain 
the following set of equations:
\begin{align}
\label{eq:28}
&I_2=\frac{1}{3},
\\
\label{eq:29}
&I_4=\frac{1}{5}+\frac{\alpha^2r_s^2}{3}
\left [-\varepsilon_c(r_s)-r_s\frac{\partial \varepsilon_c(r_s)}{\partial r_s}\right ],
\\
\label{eq:30}
&I_6=\frac{8}{105}+\frac{5}{3}I_4^2+\frac{5\alpha r_s}{9\pi}
\left [B(r_s)-\frac{2}{3}+\frac{2}{3}g(0)\right ].
\end{align}
These equations correspond to Eqs.~(\ref{eq:3})-(\ref{eq:5}), respectively, and thus 
the sum rules are now proved. Note that with use of $I_n$ and $E_{\rm F}=
1/(\alpha^2 r_s^2)$ (in Ry), we can write the average kinetic energy 
$\langle {\rm KE}\rangle$, the fluctuation of kinetic energy $\Delta {\rm KE}$, 
and $\Pi_{\rm WI}({\bm 0},0)$ as
\begin{align}
\label{eq:31}
&\langle {\rm KE} \rangle \equiv 
\langle \varepsilon_{\bm p} \rangle =3I_4\,E_{\rm F}, 
\\
\label{eq:32}
&\Delta {\rm KE} \equiv 
\sqrt{\langle (\varepsilon_{\bm p}-\langle \varepsilon_{\bm p} \rangle)^2 \rangle} 
=\sqrt{3I_6-9I_4^2}\,E_{\rm F},
\\
\label{eq:33}
&\Pi_{\rm WI}({\bm q},0) \xrightarrow[q \to 0]{}\Pi_{\rm WI}({\bm 0},0)=
\Omega_t\frac{mp_{\rm F}}{\pi^2}\,I_0.
\end{align}

\subsection{Parametrization form of $n({\bm p})$}
\label{sec:2D}

Following Gori-Giorgi and Ziesche (GZ),~\cite{Ziesche2002} we consider $n(x)$ 
in the parametrization form of
\begin{align}
\label{eq:34}
n(x)=\begin{cases}
n_0-\displaystyle{\frac{n_0-n_-}{G(0)}}\,G(x_-) &{\rm for}\ x<1,\\
\displaystyle{\frac{n_+}{G(0)}}\,G(x_+) &{\rm for}\ x>1,
\end{cases}
\end{align}
where $x_{-}$ and $x_{+}$ are, respectively, introduced as
\begin{align}
\label{eq:35}
x_{-}=&a_-(r_s)\frac{\alpha r_s}{2\pi^2}\frac{G(0)}{n_0-n_-}
\frac{1-x}{\sqrt{4\alpha r_s/\pi}}
\nonumber \\
&+b(r_s)\frac{\pi^2}{\alpha r_s}\sqrt{\frac{\pi}{3}\frac{1-\ln 2}{F''(0)}
\frac{n_0-n_-}{G(0)}}\,\frac{(1-x)^2}{x},
\end{align}
with $F''(0)/2=8.984373$ and 
\begin{align}
\label{eq:36}
x_{+}=&a_+(r_s)\frac{\alpha r_s}{2\pi^2}\frac{G(0)}{n_+}
\frac{x-1}{\sqrt{4\alpha r_s/\pi}}
\nonumber \\
&+\sqrt{\frac{3\pi (1-\ln 2)}{g(0)}
\frac{n_+}{G(0)}}\,\frac{\pi}{4\alpha r_s}\,(x-1)^4 \,F_{\infty}(x).
\end{align}
Here $G(x)$ is the Kulik function,~\cite{Kulik1961} defined as
\begin{align}
\label{eq:37}
G(x)\!=\!\int_0^{\infty}\!du \frac{R'(u)}{R(u)}
\frac{u}{u\!+\!y}
\frac{R(u)\!-\!R(y)}{u\!-\!y}\!\biggr |_{y = x/\sqrt{R(u)}},
\end{align}
with $R(u)=1-u\,\arctan(1/u)$. This function appears in the calculation of $n(x)$ 
in the random-phase approximation (RPA) and its asymptotic behavior may be 
summarized as follows: in the small-$x$ limit, 
\begin{align}
\label{eq:38}
G(x \ll 1)\!=\!G(0)\!+\!
\pi\left (\frac{\pi}{4}\!+\!\sqrt{3}\right ) x\ln x \!+\cdots,
\end{align}
with
\begin{align}
\label{eq:39}
G(0)=\int_0^{\infty}du\,\frac{-R'(u)}{R(u)}
\, \arctan \frac{1}{u}=3.353337\cdots,
\end{align}
while in the large-$x$ limit, we obtain
\begin{align}
\label{eq:40}
G(x \gg 1)=\frac{\pi}{6}(1-\ln 2)\frac{1}{x^2}\!+\cdots.
\end{align}

In accordance with this asymptotic behavior of $G(x)$, $n(x)$ in Eq.~(\ref{eq:34}) 
behaves in the following way at various limits: In the limit of $x \to 0$, 
\begin{align}
\label{eq:41}
n(x)\xrightarrow[x \to 0]{}
n_0-\frac{F''(0)}{2}\left ( \frac{\pi}{\alpha} \right )^4
\left ( \frac{\alpha r_s}{b(r_s)}\right )^2 x^2.
\end{align}
By a simple interpolation between high- and low-density limits of the coefficient of 
the $x^2$ term in Eq.~(\ref{eq:41}), GZ deduced $b(r_s)$ to be the form of
\begin{align}
\label{eq:42}
b(r_s)=(1+0.0009376925r_s^{13/4})^{1/2}.
\end{align}
We adopt this form for $b(r_s)$. In the limit of $x \to 1$, 
\begin{align}
\label{eq:43}
n(x)\xrightarrow[x \to 1+0^{\pm}]{}
n_{\pm}\,\pm \,A_{\pm}(r_s)|1-x|\ln|1-x|,
\end{align}
with $A_{\pm}(r_s)$ ``the Fermi-edge coefficient'' or the coefficient of the 
logarithmic singularity of the derivative of $n(x)$ at $x=1\pm0^+$, given by
\begin{align}
\label{eq:44}
A_{\pm}(r_s)&=\frac{1}{4}\left(\frac{\pi}{4}+\sqrt{3}\right )
\left ( \frac{\alpha r_s}{\pi} \right )^{1/2}a_{\pm}(r_s)
\nonumber \\
&\approx 0.256\sqrt{r_s}\ a_{\pm}(r_s).
\end{align}
If the electron-hole symmetry strictly holds in the electronic excitations, 
we obtain $A_-(r_s)=A_+(r_s)$, as was assumed by GZ, but at low densities in which 
electron-hole excitations are not restricted only near the Fermi surface and/or 
the dispersion of a quasi-electron is different from that of a quasi-hole, the 
symmetry will be broken in general, requiring us to determine $A_-(r_s)$ and 
$A_+(r_s)$ separately. Finally in the limit of $x \gg 1$, 
\begin{align}
\label{eq:45}
n(x)\xrightarrow[x \to \infty]{}
\frac{8}{9} \left ( \frac{\alpha r_s}{\pi} \right )^2 \frac{1}{x^8}
\frac{g(0)}{F_{\infty}(x)^2}.
\end{align}
If $F_{\infty}(x)$ approaches unity at $x \to \infty$, Eq.~(\ref{eq:45}) is reduced 
to Eq.~(\ref{eq:16}), as it should be. In fact, GZ took $F_{\infty}(x) \equiv 1$, 
but we can include the effect of the next-leading term of $O(x^{-10})$ at $x \gg 1$ 
by giving a form for $F_{\infty}(x)$ as
\begin{align}
\label{eq:46}
F_{\infty}(x)=\frac{(x-1)^2+\exp[a_{\infty}(r_s)]}{(x-1)^2+\exp[-a_{\infty}(r_s)]},
\end{align}
where $a_{\infty}(r_s)$ is an adjustable parameter related to the coefficient 
of the next-leading term. Note that $F_{\infty}(x) \equiv 1$ for $a_{\infty}(r_s) 
\equiv 0$. In actual calculations, we find that $a_{\infty}(r_s)$ plays an 
important role in fulfulling the third sum rule, Eq.~(\ref{eq:30}).

\subsection{Input data for the parametrization scheme}
\label{sec:2E}

There are six free parameters, namely, $n_0$, $n_{\pm}$, $a_{\pm}$, and $a_{\infty}$, 
in the definition of Eq.~(\ref{eq:34}). The basic idea of GZ is to determine the 
first three by the interpolation between the accurate data of $n({\bm p})$ for $1 
\le r_s \leq 5$ obtained by EPX~\cite{YT1991a} and the result in the limit of $r_s 
\to \infty$ in which GZ assume that $n({\bm p})$ is reduced to that in the Wigner 
lattice, $n_{\rm W}({\bm p})$, expressed as
\begin{align}
\label{eq:47}
n_{\rm W}({\bm p})=\frac{4\pi}{3}\left(\frac{p_{\rm F}^2}{\pi m \omega}\right )^{3/2}
\exp \left (-\frac{{\bm p}^2}{m \omega}\right ),
\end{align}
with $\omega=0.8833r_s^{-3/2}$ in a.u. In improving on the GZ scheme, we also adopt 
$n_{\rm W}({\bm p})$ as basic information for $r_s\! \gg \!10$, but the EPX data 
are upgraded by the data for $1 \!\le \!r_s\! \leq\! 10$ obtained by GW$\Gamma$ (or 
actually G$\tilde{{\rm W}}\Gamma_{\rm WI}$).~\cite{Maebashi2011} 
Then the actual parameterization formulas for $n_0(r_s)$ and $n_{\pm}(r_s)$ are 
slightly different from those in GZ. Our revised results for them are given, 
respectively, by 
\begin{align}
\label{eq:48}
n_0(r_s)&=\frac{1\!+\!t_1 r_s^{2}\!+\!t_2r_s^{5/2}}
{1\!+\!t_3 r_s^{2}\!+\!t_4r_s^{13/4}},
\\
\label{eq:49}
n_{-}(r_s)&=\frac{1\!+\!v_1 r_s\!+\!v_2r_s^{5/2}}
{1\!+\!v_3 r_s\!+\!v_4r_s^{2}\!+\!v_5r_s^{13/4}},
\\
\label{eq:50}
n_{+}(r_s)&=\frac{q_1 r_s+q_2r_s^{5/2}}
{1\!+\!q_3 r_s^{1/2}\!+\!q_4r_s^{3/2}\!+\!q_5r_s^{13/4}}.
\end{align}
with $t_1\!=\!0.013\, 813\, 294\, 1$, $t_2\!=\!0.006\, 504\, 281\, 94$, 
$t_3\!=\!0.025\, 275\, 492\, 1$, $t_4\!=\!0.001\, 015\, 523\, 77$, 
$v_1\!=\!0.198\, 200\, 080$, $v_2\!=\!0.001\, 839\, 431\, 25$, 
$v_3\!=\!0.286\, 719\, 080$, $v_4\!=\!0.013\, 899\, 539\, 9$, 
$v_5\!=\!0.000\, 473\, 501\, 718$, $q_1\!=\!0.088\, 519\, 000$, 
$q_2\!=\!0.002\, 142\, 766\, 78$, $q_3\!=\!0.387\, 350\, 521$, 
$q_4\!=\!0.079\, 971\, 845\, 3$, and $q_5\!=\!0.000\, 551\, 585\, 578$. 

\begin{figure}[htbp]
\begin{center}
\includegraphics[scale=0.46,keepaspectratio]{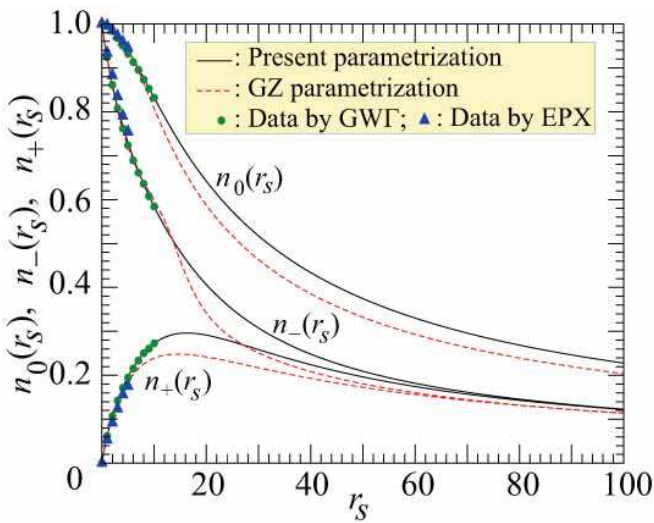}
\end{center}
\caption[Fig.3]{(Color online) Parametrized $n_0$, $n_-$, and $n_+$ as a function 
of $r_s$ in both present (solid curves) and Gori-Giorgi and Ziesche (dashed curves) 
schemes. The input data obtained by EPX (triangles) and GW$\Gamma$ (solid circles) 
are also shown.}
\label{fig:3}
\end{figure}

The obtained $n_0(r_s)$ and $n_{\pm}(r_s)$ are plotted in Fig.~\ref{fig:3}. We see 
that our results are about the same as those in GZ for $r_s <10$, but they are 
considerably different, especially for the case of $n_{+}(r_s)$; ours change much 
more smoothly with the increase of $r_s$ than those in GZ and eliminate the strange 
behavior of $n_-(r_s)$ at $r_s \approx 16$ in GZ. 

\begin{figure}[htbp]
\begin{center}
\includegraphics[scale=0.38,keepaspectratio]{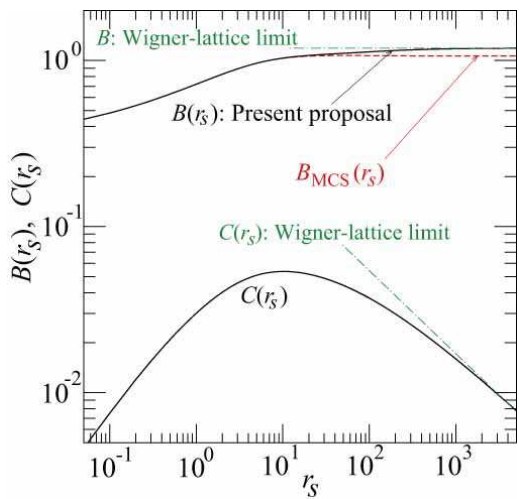}
\end{center}
\caption[Fig.4]{(Color online) Parametrized $B(r_s)$ and $C(r_s)$ as a function 
of $r_s$. The former is compared with the one proposed by Moroni, Ceperley, and 
Senatore, together with the value in the Wigner-lattice limit.}
\label{fig:4}
\end{figure}

Incidentally, if we assume that $n({\bm p})$ converges to $n_{\rm W}({\bm p})$ 
in the low-density limit, we should assess this assumption by checking the three 
sum rules with this $n_{\rm W}({\bm p})$ at $r_s \!\to\! \infty$. In fact, the first 
two sum rules are easily found to be satisfied, as already mentioned in GZ, but a 
problem exists on the third one; this $n_{\rm W}({\bm p})$ gives 
$I_4\!=\!1/(2\varepsilon_{\rm W})$ and $I_6\!=\!5/(4\varepsilon_{\rm W}^2)$ with 
$\varepsilon_{\rm W}\!=\!p_{\rm F}^2/(m\omega)\!=\!4.1692/\sqrt{r_s}$. By substituting 
those $I_4$ and $I_6$ into Eq.~(\ref{eq:30}), we obtain $\lim_{r_s \to \infty}B(r_s)
\!=\!1.1868$, while $\lim_{r_s \to \infty}B_{\rm MCS}(r_s)\!=\!1.064$, indicating the 
difference in about 10\%. Because $B_{\rm MCS}(r_s)$ was determined with the input data 
only for $r_s\!\leq \!10$, we consider it more appropriate to amend $B(r_s)$ slightly 
from $B_{\rm MCS}(r_s)$ into a form fulfilling the asymptotic value of 1.1868 as 
\begin{align}
\label{eq:51}
B(r_s)=\frac{1+[a_1+(a_2+a_3r_s)r_s]\sqrt{r_s}}
{3+[b_1+(b_2+b_3 r_s)r_s]\sqrt{r_s}},
\end{align}
with $a_1=2.161$, $a_2=0.4599$, $a_3=0.006679$, $b_1=1.594$, $b_2=0.4388$, 
and $b_3=0.005627$. In Fig.~\ref{fig:4}, this modified $B(r_s)$ is plotted 
in comparison with $B_{\rm MCS}(r_s)$ to find that they are essentially the same 
for $r_s \leq 10$. The function $C(r_s)$ in Eq.~(\ref{eq:21}) is also shown 
to see that $C(r_s)$ correctly behaves asymptotically in the Wigner-lattice limit. 

\subsection{Calculated results}
\label{sec:2F}

\begin{figure}[htbp]
\begin{center}
\includegraphics[scale=0.37,keepaspectratio]{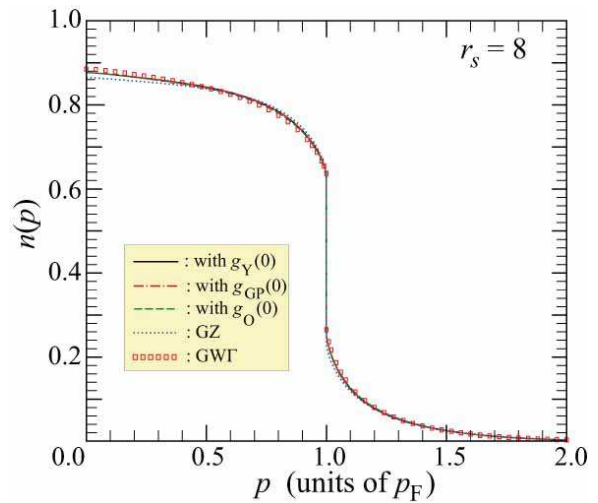}
\end{center}
\caption[Fig.5]{(Color online) An example of $n({\bm p})$ calculated with use of 
$g(0)$ in Yasuhara (solid), GP (dotted-dashed), and Overhauser (dashed) forms 
at $r_s=8$. Our results are compared with that in the GZ scheme as well as that 
in GW$\Gamma$.}
\label{fig:5}
\end{figure}

The remaining three parameters, $a_{\pm}$ (or equivalently $A_{\pm}$) and 
$a_{\infty}$, are determined so as to accurately satisfy the three sum rules 
at each $r_s$. The results for $n({\bm p})$ calculated at $r_s=8$ with use of 
three different forms, namely, Yasuhara, Overhauser, and GP, of $g(0)$ are shown 
in Fig.~\ref{fig:5}, together with those in the GZ scheme and 
GW$\Gamma$.~\cite{Maebashi2011} All these results are essentially the same, 
assuring that the present scheme works very well to obtain accurate enough 
$n({\bm p})$ satisfying three sum rules simultaneously. 

\begin{figure}[htbp]
\begin{center}
\includegraphics[scale=0.41,keepaspectratio]{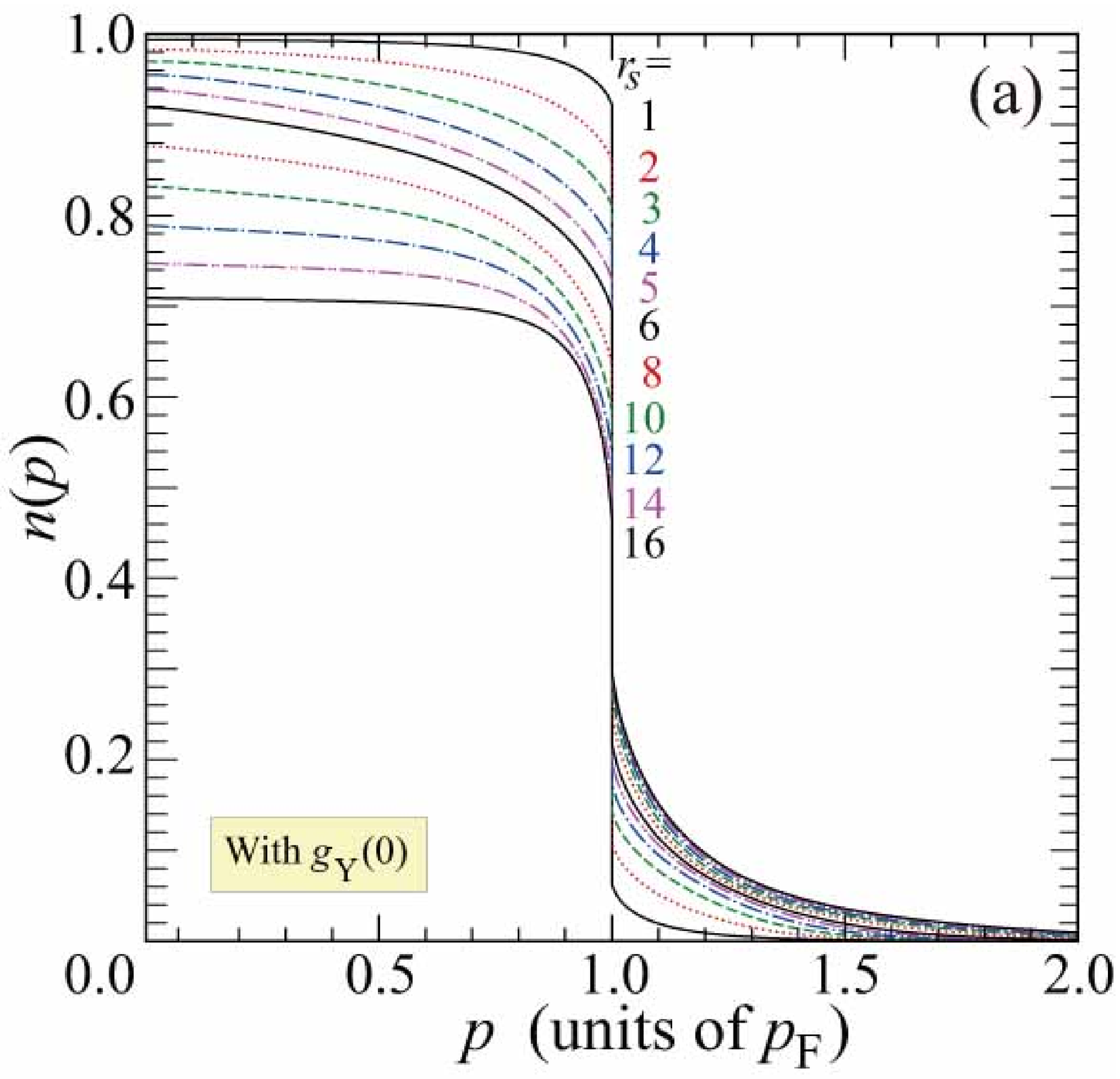}
\includegraphics[scale=0.41,keepaspectratio]{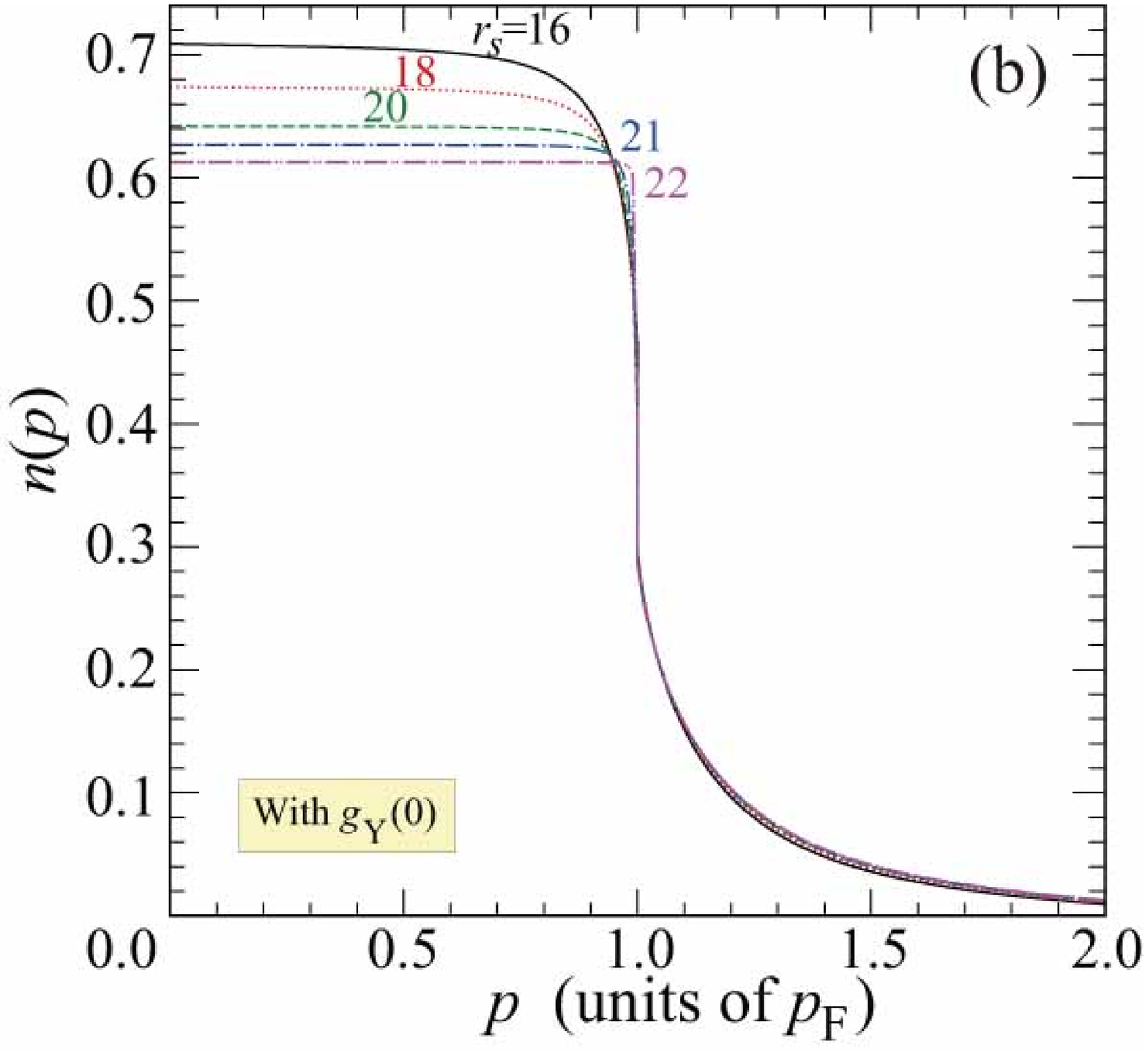}
\end{center}
\caption[Fig.6]{(Color online) Obtained $n({\bm p})$ for (a) $1\leq r_s \leq 16$ 
and (b)  $16 \leq r_s \leq 22$.}
\label{fig:6}
\end{figure}

In Fig.~\ref{fig:6}, the results of $n({\bm p})$ calculated with the choice of 
$g(0)=g_{\rm Y}(0)$ are shown for a wide range of $r_s$. In the metallic-density 
regime ($1\leq r_s \leq 5$), the obtained $n({\bm p})$ is about the same as that 
given in EPX~\cite{YT1991a} or GW$\Gamma$~\cite{YT2001}. Even in the 
dielectric-catastrophe regime, $n({\bm p})$ behaves in a normal and well-known 
manner, as long as $r_s$ is smaller than about 12. With the further increase of 
$r_s$, however, it begins to behave rather differently from the normal one 
for $|{\bm p}|<p_{\rm F}$ and eventually for $r_s>20$, it exhibits a novel feature 
in the sense that $n({\bm p})$ becomes virtually flat for $|{\bm p}|<p_{\rm F}$, 
implying that the quasi-hole excitation energy at the center of the Fermi sphere 
is about the same as that near the Fermi surface. For $|{\bm p}|>p_{\rm F}$, on 
the other hand, this anomalous flat behavior is never seen in $n({\bm p})$, 
indicating strong electron-hole asymmetric excitation spectra. Incidentally, 
a sign of this asymmetry has already been seen in the one-electron spectral 
function obtained in GW$\Gamma$ at $r_s=8$ (see, Fig. 3(b) in 
Ref.~\onlinecite{Maebashi2011}) and here we find that this tendency of asymmetry 
becomes so enhanced for $r_s> 20$ to provide the very characteristic flat 
behavior of $n({\bm p})$ in Fig~\ref{fig:6}(b).

\begin{figure}[htbp]
\begin{center}
\includegraphics[scale=0.39,keepaspectratio]{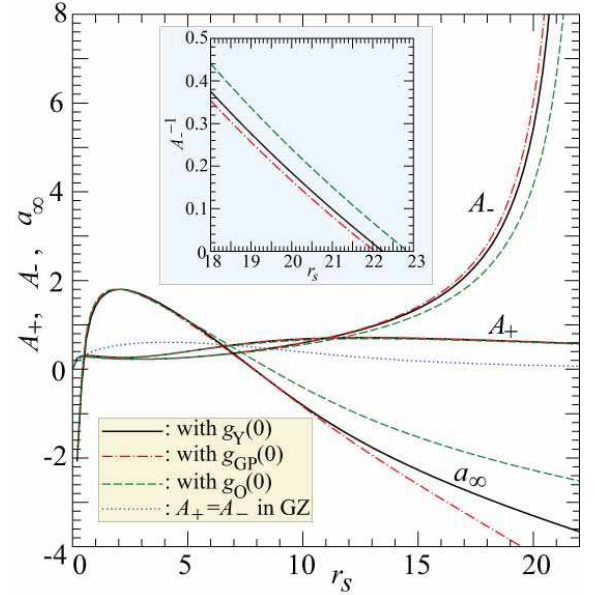}
\end{center}
\caption[Fig.7]{(Color online) Parameters $A_{\pm}$ and $a_{\infty}$ determined 
so as to satisfy the three sum rules for $n({\bm p})$ at each $r_s$. Three different 
forms for $g(0)$ are employed. For comparison, $A_{\pm}$ in GZ are also shown. 
(In GZ, $a_{\infty}=0$.) The inset displays $A_{-}^{-1}$ for the region of $r_s$ 
in which $A_{-}$ is very large.}
\label{fig:7}
\end{figure}

In Fig.~\ref{fig:7}, we plot the results for $A_{\pm}$ and $a_{\infty}$ 
determined so as to satisfy the three sum rules, Eqs.~(\ref{eq:28})-(\ref{eq:30}), 
at each $r_s$. Three different forms for $g(0)$ are employed, but the obtained 
$A_{\pm}$ is virtually independent of its choice; the difference in $g(0)$ is 
mostly compensated by the difference in $a_{\infty}$. The difference 
between $A_+$ and $A_-$ indicates the degree of electron-hole asymmetry near 
the Fermi surface; for $r_s<12$, they are essentially the same, but for 
$r_s\geq 12$, $A_-$ rapidly increases, while $A_+$ does not change much. This 
asymmetry is probably the physical reason why the GZ scheme, in which $A_-=A_+$ 
was assumed, could not give a convergent result of $n({\bm p})$ for $r_s>12$. 

\begin{figure}[htbp]
\begin{center}
\includegraphics[scale=0.39,keepaspectratio]{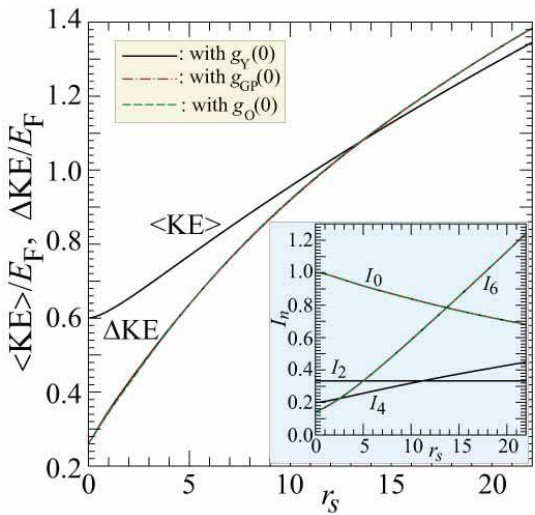}
\end{center}
\caption[Fig.8]{(Color online) The average kinetic energy and the fluctuation of 
the kinetic energy as a function of $r_s$. No difference is seen by the different 
choice of $g(0)$ because of the fulfillment of the three sum rules. 
Inset: Corresponding values of $I_n$ for $n=0, 2, 4$, and 6.}
\label{fig:8}
\end{figure}

In the parametrization scheme in Eq.~(\ref{eq:34}), the behavior of $n({\bm p})$ 
for $|{\bm p}|<p_{\rm F}$ is controlled only by the parameter $A_-$ and thus 
its anomalous flat feature is directly connected with the rapid increase of 
$A_-$ for $r_s>12$. As shown in the inset of Fig.~\ref{fig:7}, $A_-$ actually 
diverges at $r_s=\bar{r}_s^c=22.4\pm 0.4$, the error of which is deduced from the 
difference in $\bar{r}_s^c$ caused by the different choice of $g(0)$. 
For $r_s$ beyond $\bar{r}_s^c$, our scheme does not work and fails to determine 
the three parameters, $A_{\pm}$ and $a_{\infty}$, simultaneously satisfying three 
sum rules. We might be able to interpret this divergence of the Fermi-edge 
coefficient or the nonanalytic behavior at the Fermi surface as an indication of 
the occurrence of some non-Fermi-liquid phase, but at present, we do not know 
exactly whether this is the case or not, mainly because we cannot rule out the 
possibility of inadequacy of the postulated form in Eq.~(\ref{eq:34}) for the 
case of $A_- \! \gg \!1$ (or $r_s>20$). In any case, for $r_s \agt 12$, it is 
certain that some unusual situation due to electron correlation begins to appear, 
as we can see in Fig.~\ref{fig:8} in which the fluctuation of the kinetic energy 
$\Delta {\rm KE}$ becomes larger than the average one $\langle {\rm KE} \rangle$. 

\section{Dynamical Structure Factor}
\label{sec:3}

\subsection{Basic equation to determine $S({\bm q},\omega)$}
\label{sec:3A}

At $T=0$, with use of Eq.~(\ref{eq:2}) for $Q_c^R({\bm q},\omega)$, 
$S({\bm q},\omega)$ can be cast into the form of 
\begin{align}
S({\bm q},\omega)=-\frac{1}{\pi} \frac{1}{V({\bm q})}
{\rm Im}\left [\frac{1}{\varepsilon^R({\bm q},\omega)} \right ],
\label{eq:52}
\end{align}
with the retarded dielectric function $\varepsilon^R({\bm q},\omega)$, given by
\begin{align}
\varepsilon^R({\bm q},\omega)&=1+V({\bm q})\Pi^R({\bm q},\omega)
\nonumber \\
&=1+V({\bm q})\frac{\Pi_{\rm WI}^R({\bm q},\omega)}
{1-G_s^R({\bm q},\omega)V({\bm q})\Pi_{\rm WI}^R({\bm q},\omega)}.
\label{eq:53}
\end{align}
Here Eq.~(\ref{eq:6}) is employed to write $\Pi^R({\bm q},\omega)$ in terms of 
$\Pi_{\rm WI}^R({\bm q},\omega)$ and $G_s^R({\bm q},\omega)$, the latter of  which 
is given by the analytic continuation of $G_s({\bm q},i\omega)$ to the real axis 
in the complex upper $\omega$-plane ($i\omega \!\to\! \omega\!+\!i0^+$). Thus our 
task to calculate $S({\bm q},\omega)$ begins with obtaining both 
$\Pi_{\rm WI}^R({\bm q},\omega)$ and $G_s^R({\bm q},\omega)$. 

\subsection{Polarization function $\Pi_{\rm WI}^R({\bm q},\omega)$ }
\label{sec:3B}

The function $\Pi_{\rm WI}({\bm q},i\omega)$ in Eq.~(\ref{eq:7}) is rewritten as 
\begin{align}
\Pi_{\rm WI}({\bm q},i\omega)&=2\sum_{{\bm p}\sigma}n({\bm p})
\frac{\varepsilon_{{\bm p}+{\bm q}}-\varepsilon_{{\bm p}}}
{\omega^2+(\varepsilon_{{\bm p}+{\bm q}}-\varepsilon_{{\bm p}})^2}
\nonumber \\
&=\Omega_t\frac{mp_{\rm F}}{2\pi^2}P_{\rm WI}(z,iu),
\label{eq:54}
\end{align}
with the function $P_{\rm WI}(z,iu)$, introduced as 
\begin{align}
P_{\rm WI}(z,iu)
=\frac{1}{2z}\int_0^{\infty}\!xdx\,n(x)
\ln \left [ \frac{u^2+(x+z)^2}{u^2+(x-z)^2}\right ],
\label{eq:55}
\end{align}
where $x=|{\bm p}|/p_{\rm F}$, $z=|{\bm q}|/2p_{\rm F}$, and $u=(\omega/4E_{\rm F})
/z$. By analytic continuation, $\Pi_{\rm WI}^R({\bm q},\omega)$ is obtained as 
\begin{align}
\Pi_{\rm WI}^R({\bm q},\omega)
=\Omega_t \frac{mp_{\rm F}}{2\pi^2}P_{\rm WI}(z,u\!+\!i0^+), 
\label{eq:56}
\end{align}
with $P_{\rm WI}(z,u\!+\!i0^+)$, given by
\begin{align}
P_{\rm WI}(z,u\!+\!i0^+)\!=\!&
\frac{1}{2z}\!\int_0^{\infty}\!xdx\, n(x)\!
\Bigl \{ \!\ln[u^2\!-\!(x\!+\!z)^2\!+\!i0^+] 
\nonumber \\
&-\!\ln[u^2\!-\!(x\!-\!z)^2\!+\!i0^+]\!\Bigr \},
\label{eq:57}
\end{align}
where $\ln(\omega)$ is so defined as $\ln(\omega)\!=\!\ln(|\omega|)\!+\!
i{\rm arg}(\omega)$ with arg($\omega$) the argument of $\omega$ in the upper 
$\omega$-plane being between 0 and $\pi$. In performing the numerical evaluation of 
this integral over the variable $x$, care must be exerted at both $x\!=\!|u\!-\!z|$ 
and $x\!=\!u\!+\!z$ in order to accurately obtain the principal values of the integral.

\begin{figure}[hbtp]
\begin{center}
\includegraphics[scale=0.55,keepaspectratio]{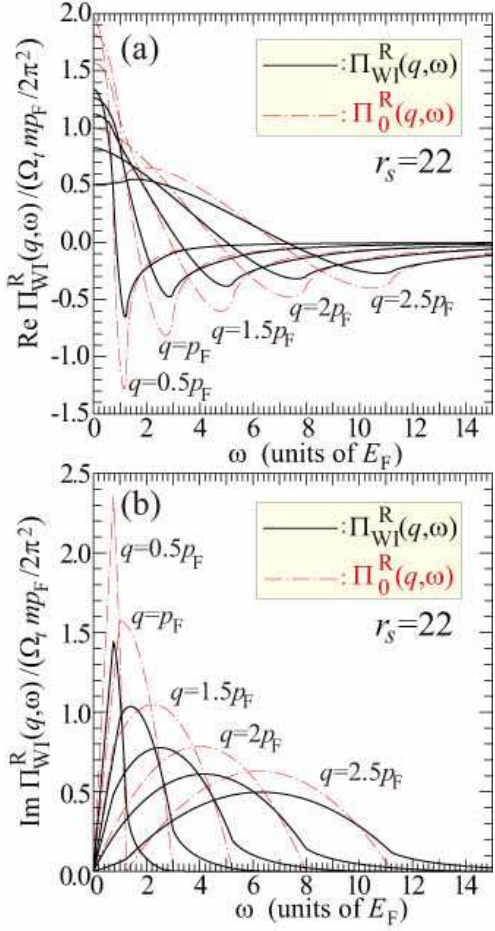}
\end{center}
\caption[Fig.9]{(Color online) The retarded polarization function 
$\Pi_{\rm WI}^R({\bm q},\omega)$ in comparison with $\Pi_{0}^R({\bm q},\omega)$ 
at $r_s=22$ for (a) real and (b) imaginary parts.}
\label{fig:9}
\end{figure}

With $n({\bm p}) [=n(x)]$ obtained in Sec.~\ref{sec:2F}, we can calculate 
$\Pi_{\rm WI}^R({\bm q},\omega)$ through Eqs.~(\ref{eq:56}) and (\ref{eq:57}). 
The overall structure of the calculated $\Pi_{\rm WI}^R({\bm q},\omega)$ does not 
depend much on $r_s$, in spite of the apperance of an anomalous feature in 
$n({\bm p})$ for $r_s>20$; as shown in Fig.~\ref{fig:9}, even at $r_s=22$, the 
results of $\Pi_{\rm WI}^R({\bm q},\omega)$  given as a function of $\omega$ 
at various values of ${\bm q}$ (the solid curves) retain the main features of 
those of $\Pi_{0}^R({\bm q},\omega)$ the Lindhard function in the retarded form 
(the dotted-dashed curves). We can understand this insensitivity of 
$\Pi_{\rm WI}^R({\bm q},\omega)$ to the anomalous feature in $n({\bm p})$ by 
paying attention to the fact that $\Pi_{\rm WI}^R({\bm q},\omega)$ is mostly 
controlled by the sum rules for $n({\bm p})$; the details of $n({\bm p})$ are 
smeared out by the integral over the variable $x$ in Eq.~(\ref{eq:57}). 

Note, however, that there is a very important difference between 
$\Pi_{0}^R({\bm q},\omega)$ and $\Pi_{\rm WI}^R({\bm q},\omega)$; 
${\rm Im}\Pi_{0}^R({\bm q},\omega)$ does not vanish only in the electron-hole 
single-pair excitation region in $({\bm q},\omega)$ space, while basically 
${\rm Im}\Pi_{\rm WI}^R({\bm q},\omega)$ is not zero but positive definite for 
$\omega>0$ at any ${\bm q}$ due to the inclusion of the multiple electron-hole pair 
excitations by the use of $n({\bm p})$ instead of $n_0({\bm p})$. For this reason, 
as long as Eq.~(\ref{eq:53}) is employed, we do not need to introduce an artificial 
finite broadening width $\gamma$ in the numerical calculation of the dielectric 
function, in sharp contrast with the usual treatment of $\omega \to \omega+i\gamma$ 
in the calculation of the plasmon-peak structure in $S({\bm q},\omega)$ with using 
$\Pi_{0}^R({\bm q},\omega)$ or its simple extension in which no account is taken of 
the multiple electron-hole pair excitations. 

\subsection{Dynamical local-field factor $G_s^R({\bm q},\omega)$}
\label{sec:3C}

There is a long history of researches on the dynamical local-field factor and its 
parametrization.~\cite{Gross1985,Dabrowski1986,Iwamoto1987,Holas1989,Nifosi1998,
Lein2000,Qian2002,Morawetz2002,Constantin2007} Those studies are mainly motivated by 
the construction of the exchange-correlation kernel 
$f_{\rm xc}({\bm r},{\bm r'};\omega)$ in the time-dependent density functional 
theory (TDDFT) and thus they are concerned with either $G_+({\bm q},i\omega)$ or 
$G_+^R({\bm q},\omega)$, but not with $G_s({\bm q},i\omega)$. So far a 
parametrization form for $G_s({\bm q},i\omega)$ is given only by 
RA~\cite{Richardson1994}, proposing 
\begin{align}
G_s({\bm q},i\omega)
=\frac{a_s(i\Omega)z^2+2[1-g(0)]b_s(i\Omega)z^8/3}
{1+c_s(i\Omega)z^2+b_s(i\Omega)z^8},
\label{eq:58}
\end{align}
with $\Omega\!\equiv\!zu\!=\!\omega/4E_{\rm F}$. This form of 
$G_s({\bm q},i\omega)$ satisfies the rigorous limit due to 
Niklasson~\cite{Niklasson1974} at $|{\bm q}| \to \infty$ as $G_s({\bm q},i\omega) 
\to 2[1-g(0)]/3$, irrespective of $\omega$. 

The function $a_s(i\Omega)$ is so determined as to satisfy the constraints in 
the small-$|{\bm q}|$ regime, namely, the compressibility sum rule at $\Omega=0$ 
and the third-moment sum rule~\cite{Puff1965,Pathak1973,Iwamoto1984} at 
$\Omega \to \infty$, leading to the following form:
\begin{align}
a_s(i\Omega)=\lambda_s^{(\infty)}+
\frac{\lambda_s^{(0)}-\lambda_s^{(\infty)}}
{1+\beta_1\gamma_s\Omega+(\beta_2\gamma_s\Omega)^2},
\label{eq:59}
\end{align}
where $\lambda_s^{(0)}$ and $\lambda_s^{(\infty)}$ are, respectively, defined as
\begin{align}
\lambda_s^{(0)}&=\frac{\pi}{\alpha r_s}\left (\frac{1}{I_0}
-\frac{\kappa_F}{\kappa}\right ),
\label{eq:60}
\\
\lambda_s^{(\infty)}&=\frac{3}{5}-\frac{2}{5}\pi \alpha r_s
\left[ r_s\frac{\partial \varepsilon_c(r_s)}{\partial r_s}
+2\varepsilon_c(r_s) \right ],
\label{eq:61}
\end{align}
where the ratio of the compressibilities with and without the interactions, 
$\kappa$ and $\kappa_F$, can be calculated as
\begin{align}
\frac{\kappa_F}{\kappa}=1-\frac{\alpha r_s}{\pi}+\frac{\alpha^2r_s^3}{6}
\left [ r_s \frac{\partial^2 \varepsilon_c(r_s)}{\partial r_s^2}
-2\frac{\partial \varepsilon_c(r_s)}{\partial r_s}\right ].
\label{eq:62}
\end{align}
As for the two functions, $c_s(i\Omega)$ and $b_s(i\Omega)$, in Eq.~(\ref{eq:58}), 
by the consideration of the condition to obtain the maximum value of 
$4[1-g(0)]/3$ for $G_s({\bm q},i\omega)$ at $z^2 \approx \Omega \to \infty$, 
RA gave the following forms:
\begin{align}
c_s(i\Omega)=&\frac{3}{4}\frac{\lambda_s^{(\infty)}}{1-g(0)}
-\frac{\displaystyle{\frac{4}{3}-\frac{1}{\alpha_{\rm RA}}
+\frac{3}{4}\frac{\lambda_s^{(\infty)}}
{1-g(0)}}}
{1+\gamma_s\Omega},
\label{eq:63}
\\
b_s(i\Omega)=&\frac{a_s(i\Omega)}{3a_s(i\Omega)-2[1-g(0)]
[3c_s(i\Omega)+4/(1+\Omega)]/3}
\nonumber \\
&\times \frac{1}{1+\gamma_1\Omega+\gamma_2\Omega^2+\gamma_3\Omega^3+\Omega^4},
\label{eq:64}
\end{align}
with the parameter $\gamma_s$ determined as
\begin{align}
\gamma_s=\frac{9}{16}\frac{\lambda_s^{(\infty)}}{1-g(0)}
+1-\frac{3}{4}\frac{1}{\alpha_{\rm RA}}. 
\label{eq:65}
\end{align}
In Eqs.~(\ref{eq:59}), (\ref{eq:63}), and (\ref{eq:64}), six parameters, 
$\alpha_{\rm RA}$, $\beta_1$, $\beta_2$, $\gamma_1$, $\gamma_2$, and $\gamma_3$, 
are at our disposal, but RA made the following choice: $\alpha_{\rm RA}=0.9$, 
$\beta_1=0$, $\beta_2=1$, $\gamma_1=\gamma_3=4$, and $\gamma_2=6$. 

\begin{figure}[htbp]
\begin{center}
\includegraphics[scale=0.38,keepaspectratio]{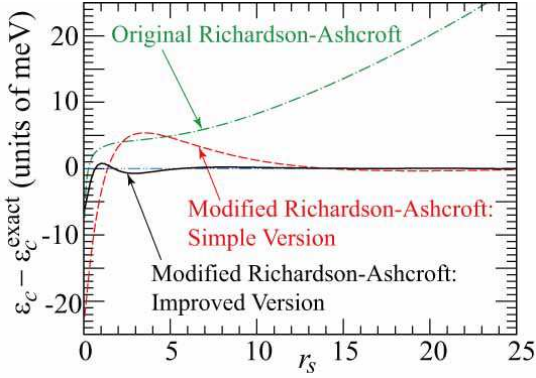}
\end{center}
\caption[Fig.10]{(Color online) Difference between the ``exact'' value of 
$\varepsilon_c$ due to Perdew and Wang and the one calculated through the adiabatic 
connection formula with using the local-field factor in the original RA (the 
dotted-dashed curve), a simple version of the modified RA (the dashed curve), and 
an improved version of the modified RA (the solid curve) schemes.}
\label{fig:10}
\end{figure}

A similar parametrization scheme was proposed by RA for $G_n({\bm q},i\omega) 
[\equiv G_+({\bm q},i\omega)\!-\!G_s({\bm q},i\omega)]$. As a matter of fact, with 
use of Eq.~(\ref{eq:24}), $G_n({\bm q},i\omega)$ is given exactly as 
\begin{align}
G_n({\bm q},i\omega) &= G_n^{\rm exact}({\bm q},i\omega) 
\nonumber \\
&\equiv 
\frac{1}{V({\bm q})}\left [ \frac{1}{\Pi_{0}({\bm q},i\omega)}
-\frac{1}{\Pi_{\rm WI}({\bm q},i\omega)}\right ].
\label{eq:66}
\end{align}
We have calculated $G_n^{\rm exact}({\bm q},i\omega)$ through Eq.~(\ref{eq:66}) 
with $\Pi_{\rm WI}({\bm q},i\omega)$ given by Eqs.~(\ref{eq:54}) and 
(\ref{eq:55}) and compared with $G_n^{\rm RA}({\bm q},i\omega)$, the 
parametrized form for $G_n({\bm q},i\omega)$ due to RA, to find that 
$G_n^{\rm RA}({\bm q},i\omega)$ is in fact very reliable; in particular, 
no appreciable difference can be seen for the quantities given by 
integrals over ${\bm q}$ and $\omega$, such as $\varepsilon_c(r_s)$ obtained 
through the adiabatic connection formula, Eq.~(27) in Ref.~\onlinecite{Lein2000}, 
in which we have calculated with use of either $G_s({\bm q},i\omega)+
G_n^{\rm RA}({\bm q},i\omega)$ or $G_s({\bm q},i\omega)+
G_n^{\rm exact}({\bm q},i\omega)$. With this assessment of the RA parametrization 
scheme, we can expect that the parametrized $G_s({\bm q},i\omega)$ will also be 
reliable. Incidentally, this RA scheme comes to be known to possess a very good 
feature~\cite{Lein2000}; it provides a quite accurate $\varepsilon_c(r_s)$ in 
reference to its ``exact'' value $\varepsilon_c^{{\rm exact}}(r_s)$ given by 
Perdew and Wang~\cite{PW1992} (see the dotted-dashed curve in Fig.~\ref{fig:10}). 

By analytic continuation ($i\omega \!\to \!\omega\!+\!i0^+$), we obtain 
$G_s^R({\bm q},\omega)$ from $G_s({\bm q},i\omega)$ in Eq.~(\ref{eq:58}), but with 
$\beta_i$s chosen in RA, $G_s^R({\bm q},\omega)$ at small ${\bm q}$ suffers from 
the existence of unphysical divergence at $\omega\!=\!\pm 4E_{\rm F}/\gamma_s$. A 
simple way to eliminate this unphysical divergence while keeping the good feature 
of providing accurate $\varepsilon_c(r_s)$ is found to choose $\alpha_{\rm RA}\!=
\!0.958/(1\!+\!0.006r_s)$, $\beta_1\!=\!0.8$, and $\beta_2\!=\!0.4$ with $\gamma_i$s 
unchanged. This simple version of the modified RA scheme gives better 
$\varepsilon_c(r_s)$ than the original RA scheme (see Fig.~\ref{fig:10}). A closer 
analysis of this scheme, however, reveals that unphysical divergence appears for 
$2p_{\rm F} \! \lesssim \!|{\bm q}| \! \lesssim 3p_{\rm F}$ with $\gamma_i$s 
chosen as those in RA.

Under these circumstances, we have made an extensive search of suitable values for 
all free parameters in Eqs.~(\ref{eq:59}), (\ref{eq:63}), and (\ref{eq:64}) with 
considering the following four criteria; (i) $G_s^R({\bm q},\omega)$ contains no 
unphysical divergence for real $\omega$, (ii) $G_s^R({\bm q},\omega)$ should be 
analytic in the upper complex $\omega$-plane, in accord with the causality 
condition,~\cite{Gross1985,Dabrowski1986} (iii) $G_s({\bm q},i\omega)$ accurately 
provides $\varepsilon_c$ in a wide range of $r_s$, and (iv) $G_s^R({\bm q},\omega)$ 
combined with $\Pi_{\rm WI}^R({\bm q},\omega)$ gives the results of 
$S({\bm q},\omega)$ in good agreement with those in GW$\Gamma$~\cite{YT2002} 
for $1 \! \leq \! r_s \! \leq \! 5$. As a result of this search, we have chosen the 
following set of parameters, which we call an improved version of the modified RA 
scheme; $\beta_1\!=\!\beta_2\!=\!0.8$, $\gamma_1\!=\!\gamma_2\!=\!\gamma_3\!=\!1$ and 
\begin{align}
\alpha_{\rm RA}(r_s)=\frac{\alpha_0}
{1+\left [\alpha_1
+\alpha_2/(1+\alpha_3r_s+\alpha_4r_s^2) \right ]r_s},
\label{eq:67}
\end{align}
with $\alpha_0\!=\!1.2110$, $\alpha_1\!=\!0.01064$, $\alpha_2\!=\!0.10296$, 
$\alpha_3\!=\!-0.2910$, and $\alpha_4\!=\!0.2540$. 

\subsection{Ghost exciton mode $\tilde{\omega}_{\rm ex}({\bm q})$}
\label{sec:3D}

With $\Pi_{\rm WI}({\bm q},i\omega)$ and $G_s({\bm q},i\omega)$ thus determined, 
we can calculate $\Pi({\bm q},i\omega)$ through Eq.~(\ref{eq:6}) to find that it 
actually diverges at $i\omega=\pm i\tilde{\omega}_{\rm ex}({\bm q})$ for $r_s\!>\!r_s^c$ 
in accord with the discussion in Ref.~\onlinecite{Takayanagi1997}. Numerically, 
$\tilde{\omega}_{\rm ex}({\bm q})$ can be determined by the search of zero of 
$\Pi({\bm q},i\omega)^{-1}$ for $\omega>0$. 

\begin{figure}[htbp]
\begin{center}
\includegraphics[scale=0.44,keepaspectratio]{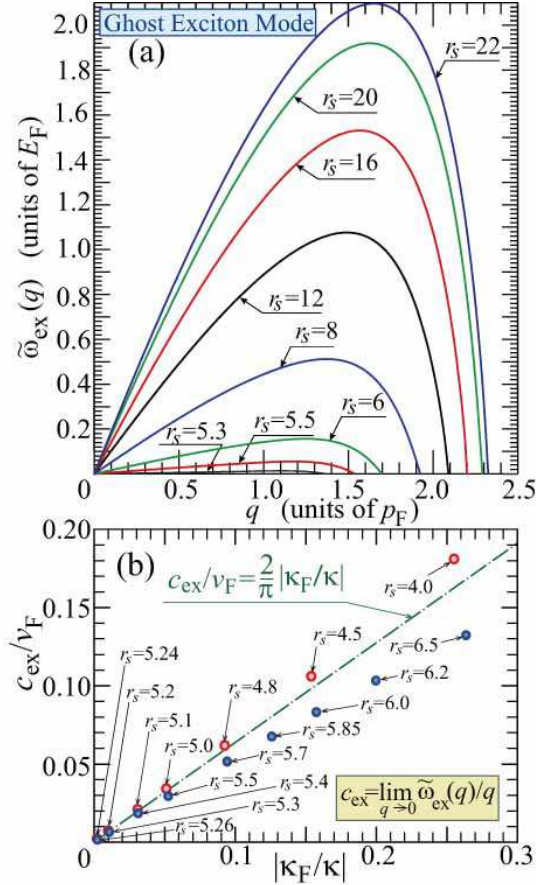}
\end{center}
\caption[Fig.11]{(Color online) (a) Dispersion relation of the ghost exciton mode 
at various values of $r_s$.(b) Sound velocity of the ghost exciton 
mode in units of the Fermi velocity $v_{\rm F}$ plotted as a function of 
$|\kappa_F/\kappa|$.}
\label{fig:11}
\end{figure}

In Fig.~\ref{fig:11}(a), the obtained results of $\tilde{\omega}_{\rm ex}({\bm q})$ 
are shown as a function of $q (=|{\bm q}|)$ at several values of $r_s$ in the 
dielectric-catastrophe regime of $r_s>r_s^c$. For small $q$, 
$\tilde{\omega}_{\rm ex}({\bm q})$ increases in proportion to $q$, just like a sound 
mode, and ``the sound velocity'' $c_{\rm ex}$ changes in proportion to 
$|\kappa_F/\kappa|$ with the coefficient $2v_{\rm F}/\pi$ ($v_{\rm F}$: the Fermi 
velocity) for $r_s$ near $r_s^c$, as shown in Fig.~\ref{fig:11}(b). Note that this 
behavior of $c_{\rm ex}$ is quite similar to that for ``the exciton mode'' 
identified by the peak position of ${\rm Im}\,\varepsilon^R({\bm q},\omega)$ for 
$r_s<r_s^c$.~\cite{YT2005} Thus it is more appropriate to call the mode of 
$\tilde{\omega}_{\rm ex}({\bm q})$ on the imaginary $\omega$ axis ``the ghost 
exciton mode'' than ``the ghost plasmon'' which was suggested in 
Ref.~\onlinecite{Takayanagi1997}.

\begin{figure}[htbp]
\begin{center}
\includegraphics[scale=0.36,keepaspectratio]{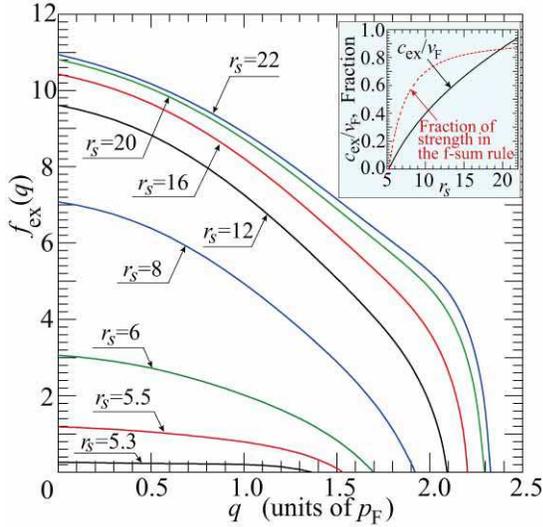}
\end{center}
\caption[Fig.12]{(Color online) Oscillator strength of the ghost exciton mode. 
Inset: Fraction of this oscillator strength to the full strengths in the limit of 
$q \to 0$, together with the sound velocity in units of $v_{\rm F}$, plotted as 
a function of $r_s$.}
\label{fig:12}
\end{figure}

The oscillator strength $f_{\rm ex}({\bm q})$ in the definition of 
$\Pi_a({\bm q},i\omega)$ in Eq.~(\ref{eq:1}) is determined by the evaluation of 
$[\omega-\tilde{\omega}_{\rm ex}({\bm q})]\Pi({\bm q},i\omega)$ in the limit of 
$\omega \to \tilde{\omega}_{\rm ex}({\bm q})$ and the obtained results are shown in 
Fig.~\ref{fig:12}. Incidentally, the asymptotic behavior of $\Pi({\bm q},i\omega)$ 
in Eq.~(\ref{eq:6}) is found to be 
\begin{align}
\Pi({\bm q},i\omega) 
\xrightarrow[\omega \to \infty]{}
\frac{4\pi}{V({\bm q})}\,
\frac{ne^2}{m}\,\frac{1}{\omega^2},
\label{eq:68}
\end{align}
from which we can prove the f-sum rule for $S({\bm q},\omega)$ as
\begin{align}
\int_{-\infty}^{\infty}d\omega\,\omega S({\bm q},\omega)
= \frac{N{\bm q}^2}{2m}.
\label{eq:69}
\end{align}
On the other hand, $\Pi_a({\bm q},i\omega)$ behaves as 
\begin{align}
\Pi_a({\bm q},i\omega) 
\xrightarrow[\omega \to \infty]{}
\frac{1}{V({\bm q})}\frac{ne^2}{m}
\frac{f_{\rm ex}({\bm q})}{\omega^2}. 
\label{eq:70}
\end{align}
Comparison of Eq.~(\ref{eq:70}) with Eq.~(\ref{eq:68}) indicates that the ghost 
exciton mode contributes to the full oscillator strengths with the fraction of 
$f_{\rm ex}({\bm q})/4\pi$, which becomes more than 80\% at $q \to 0$ for $r_s>14$. 
Thus, in such a dilute electron gas, we come to notice that not the plasmon but 
the exciton mode will play a main role in the whole excitation spectra. 

\subsection{Calculated results of $S({\bm q},\omega)$}
\label{sec:3E}

\begin{figure}[htbp]
\begin{center}
\includegraphics[scale=0.43,keepaspectratio]{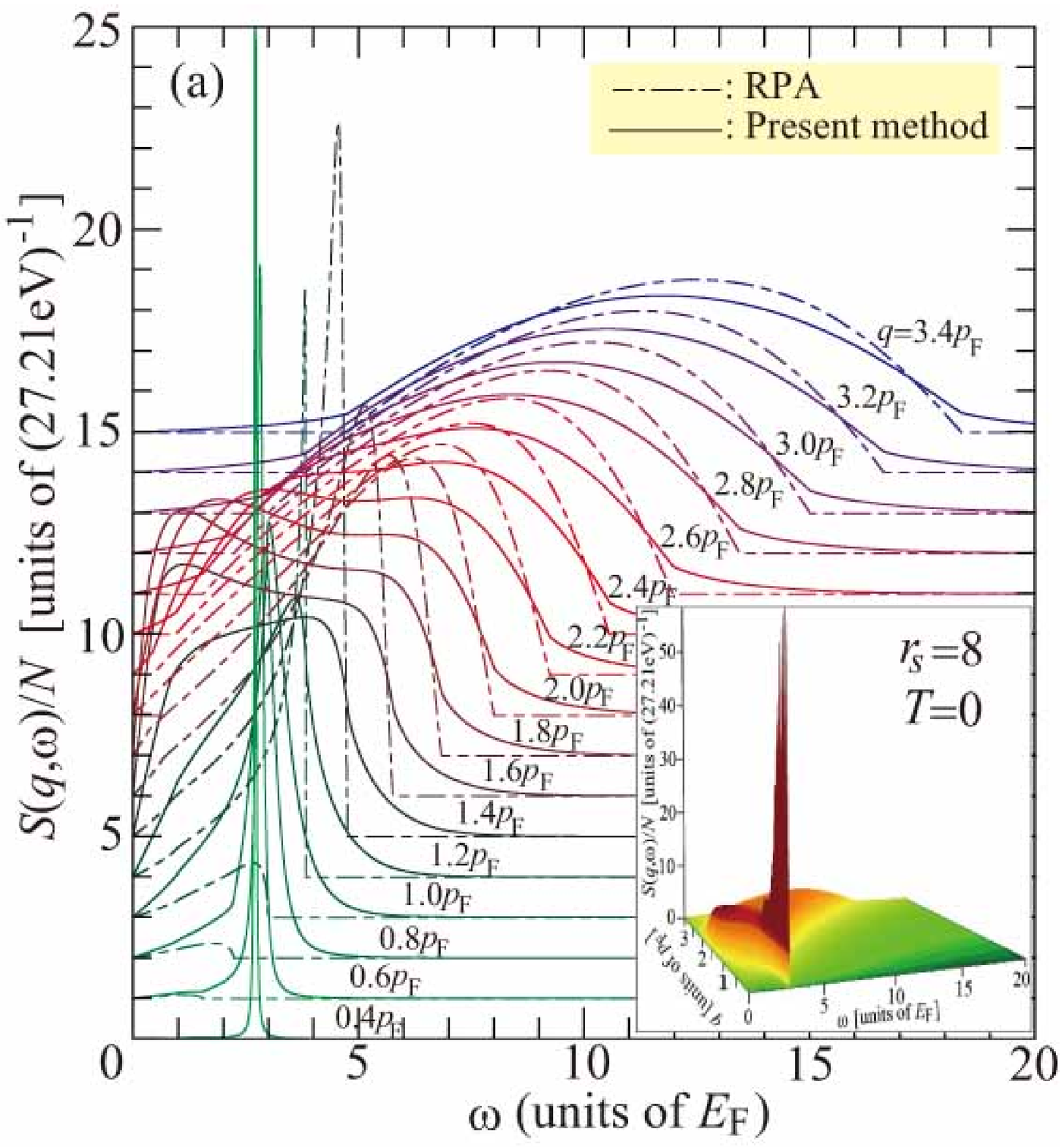}
\includegraphics[scale=0.43,keepaspectratio]{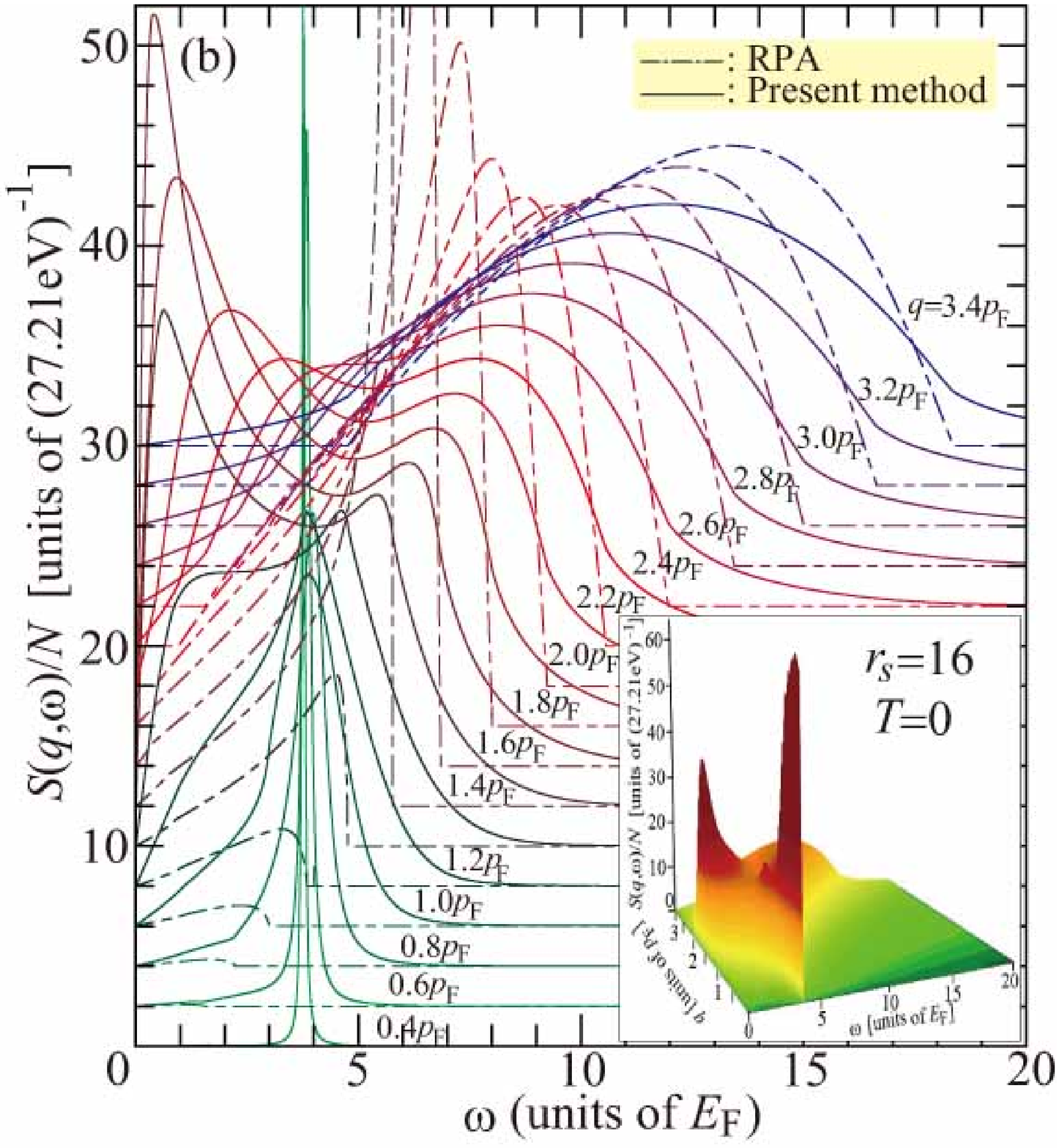}
\end{center}
\caption[Fig.13]{(Color online) The Dynamical structure factor at (a) $r_s=8$ 
and (b) $r_s=16$. Inset: Bird's-eye view.}
\label{fig:13}
\end{figure}

With $\Pi_{\rm WI}^R({\bm q},\omega)$ and $G_s^R({\bm q},\omega)$ already given, 
we can calculate $S({\bm q},\omega)$ through Eqs.~(\ref{eq:52}) and (\ref{eq:53}). 
As mentioned before, in the usual metallic region ($1 \leq r_s \leq 5$), the form of 
$G_s^R({\bm q},\omega)$ is finely tuned to reproduce the results of 
$S({\bm q},\omega)$ that are already obtained in Ref.~\onlinecite{YT2002}. 
As illustrated in Fig.~\ref{fig:1}(a) for $r_s=4$, the structure of 
$S({\bm q},\omega)$ is featured by the single plasmon peak and the exciton 
contribution appears not as a peak but only as a shoulder structure in the 
low-$\omega$ region. (To be more definite, see the structure ``$a$'' specified 
in Fig. 1 in Ref.~\onlinecite{YT2002}.) 

Even in the dielectric-catastrophe regime ($r_s>r_s^c$), the main feature of 
$S({\bm q},\omega)$ hardly changes at $r_s$ near $r_s^c$, but as $r_s$ increases 
further, the shoulder structure due to the excitonic effect gradually evolves into a 
broad peak, as illustrated in Fig.~\ref{fig:13}(a) at $r_s=8$. The peak becomes 
sharper with the further increase of $r_s$ and appears as a clear peak for $r_s 
\approx 10$. Eventually for $r_s \agt 14$, combined with the plasmon peak, 
$S({\bm q},\omega)$ is characterized by a twin-peak structure, as shown at 
$r_s=16$ in Fig.~\ref{fig:13}(b). At $r_s=20$ or larger, ``the excitonic peak'' grows 
even bigger than the plasmon one [see Fig.~\ref{fig:1}(b)]. 

\begin{figure}[htbp]
\begin{center}
\includegraphics[scale=0.44,keepaspectratio]{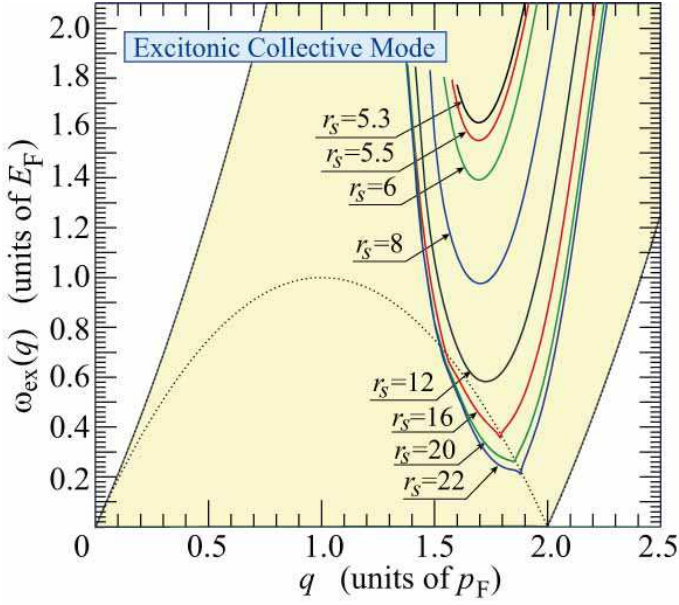}
\end{center}
\caption[Fig.14]{(Color online) Dispersion relation of the excitonic collective 
mode as determined from the peak position in $S({\bm q},\omega)$ in $(q,\omega)$ 
space. The shaded area represents the electron-hole single-pair excitation region 
and the dotted curve indicates the boundary of $\omega=(2p_{\rm F}q-q^2)/2m$. }
\label{fig:14}
\end{figure}

In Fig.~\ref{fig:14}, we have plotted $\omega_{\rm ex}(q)$ the excitonic peak position 
(including the shoulder position, if a peak structure is not well identified) 
at each $q$ in $(q,\omega)$ space. This plot shows that $\omega_{\rm ex}(q)$ is 
always inside the single-pair excitation region, as it should be for the excitonic 
effect working on an electron-hole single-pair excitation. 

An even more interesting fact is that if $\omega_{\rm ex}(q)$ is in the region of 
$\omega \leq (2p_{\rm F}q-q^2)/2m$, the corresponding peak in $S({\bm q},\omega)$ 
is very sharp. In particular, $\omega_{\rm ex}(q)$ becomes lowest if it lies just 
on this boundary and the highest peak in $S({\bm q},\omega)$ appears in its very 
vicinity. Furthermore, with the increase of $r_s$, this lowest-energy peak position 
$(q_{\rm min},\omega_{\rm ex}(q_{\rm min}))$ becomes lower approximately in 
proportion to $r_s$ along the boundary curve, so that with the further increase of 
$r_s$ up to about 30, we may imagin a critical situation of $q_{\rm min} \to 
2p_{\rm F}$ and $\omega_{\rm ex}(q_{\rm min}) \to 0$ by extrapolation. If this 
critical situation were actually realized, we might expect the occurrence of a very 
exotic phase transition brought about by the spontaneous excitation of a macroscopic 
number of excitons or ``the exciton condensation''. 

\subsection{Retarded dielectric function $\varepsilon^R({\bm q},\omega)$}
\label{sec:3F}

Let us make a detailed analysis of the retarded dielectric function 
$\varepsilon^R({\bm q},\omega)$ in order to investigate the relation between the 
excitonic-peak position (or the excitonic collective mode) $\omega_{\rm ex}(q)$ 
and the ghost exciton mode $\tilde{\omega}_{\rm ex}({\bm q})$. 

\begin{figure}[htbp]
\begin{center}
\includegraphics[scale=0.46,keepaspectratio]{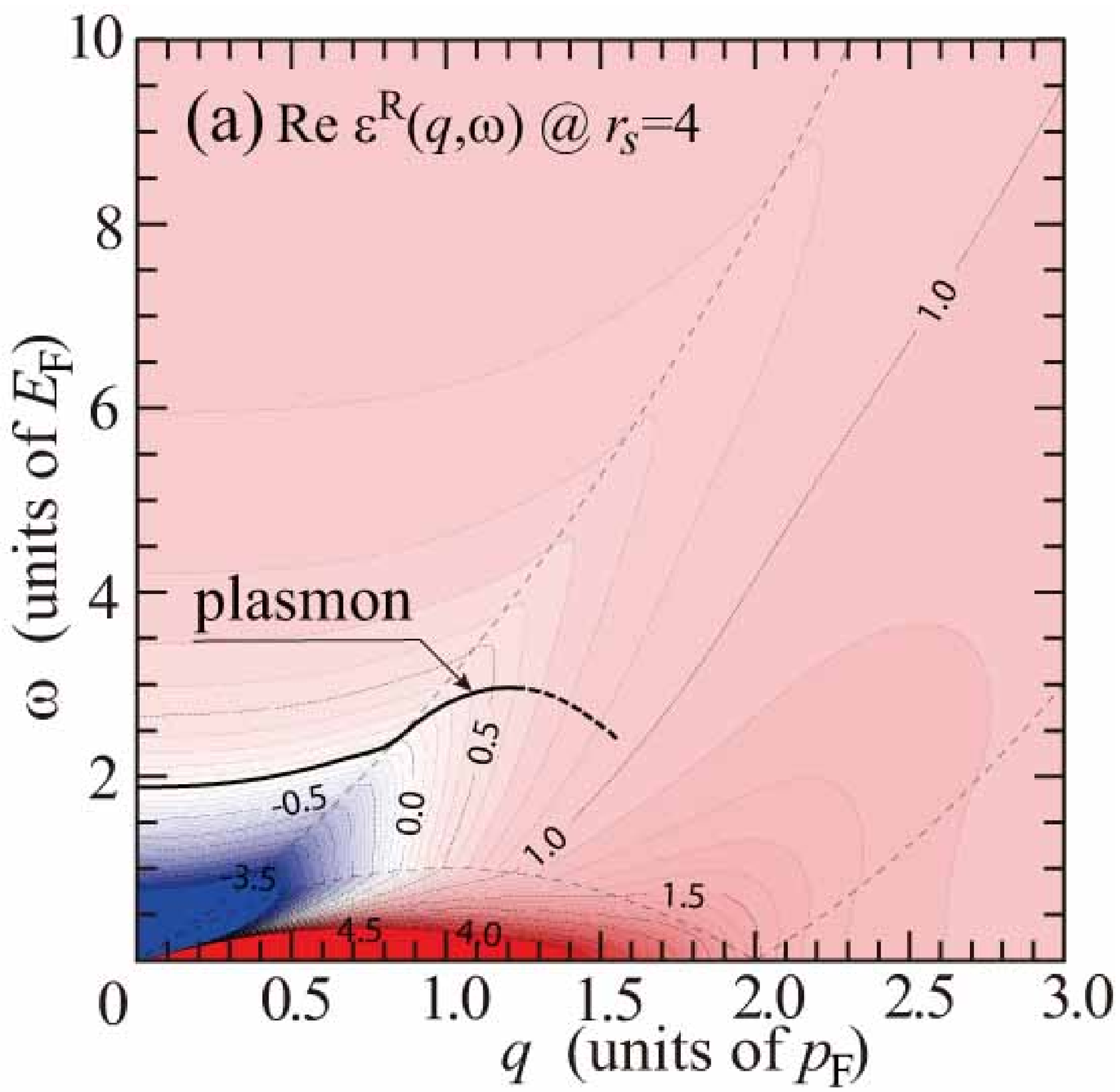}
\includegraphics[scale=0.46,keepaspectratio]{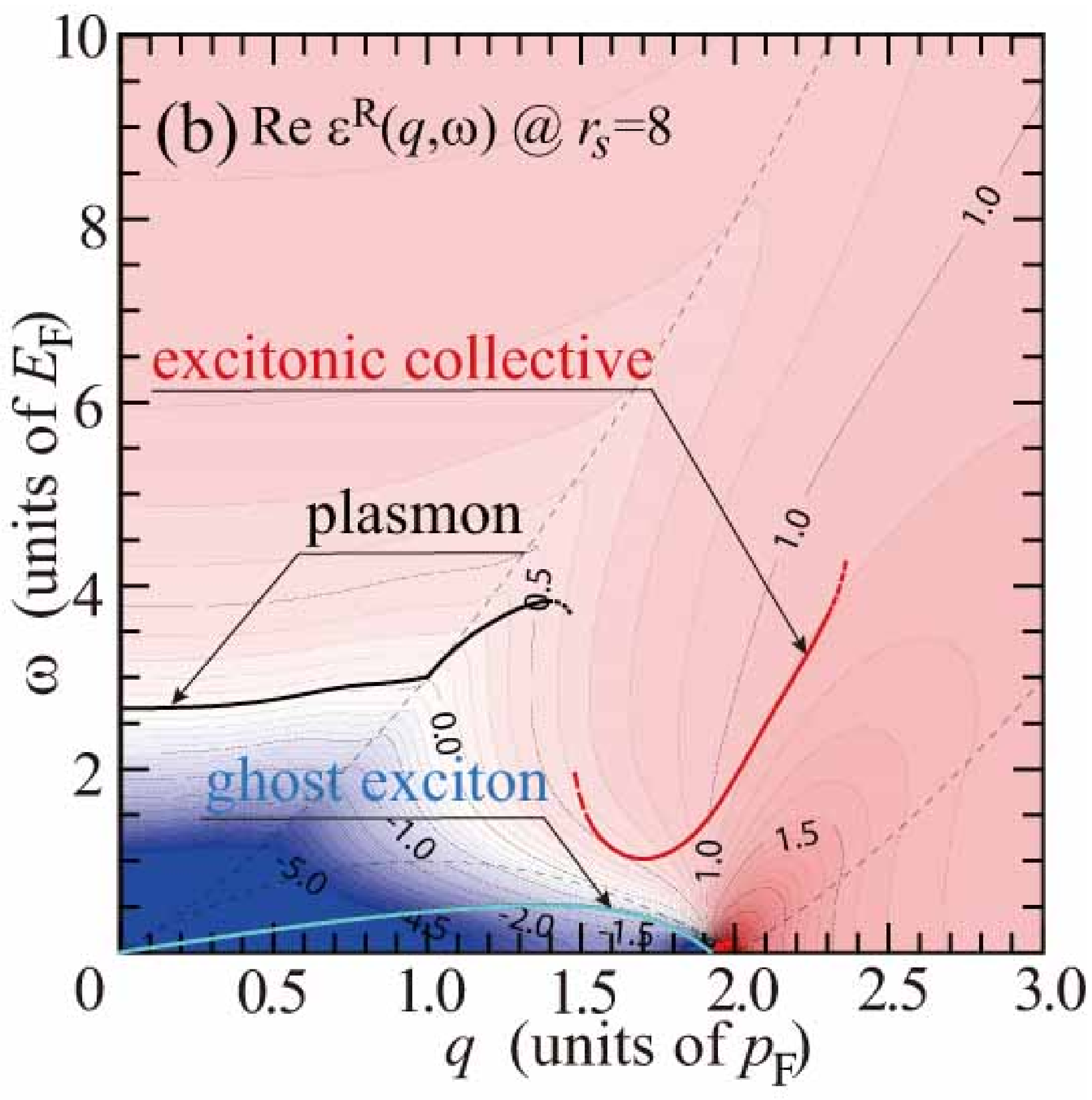}
\end{center}
\caption[Fig.15]{(Color online) Two-dimensional contour map of the real part of 
$\varepsilon^R({\bm q},\omega)$ at (a) $r_s=4$ and (b) $r_s=8$.}
\label{fig:15}
\end{figure}

In the conventional metals without the dielectric catastrophe as represented by the 
electron gas at $r_s=4$, the overall behavior of $\varepsilon^R({\bm q},\omega)$ is 
well known and rather simple; as shown in Fig.~\ref{fig:15}(a) for its real part 
in $(q,\omega)$ space, there is only a single singular point at $q=\omega=0$ in this 
function, characterized by $\varepsilon^R({\bm q},\omega) \approx 1-\omega_{pl}^2/
\omega^2$ in the limit of $\omega \to 0$ with keeping $v_{\rm F}q/\omega \ll 1$ (the 
$\omega$-limit) and $\varepsilon^R({\bm q},\omega) \approx 1+(\kappa/\kappa_F) 
q_{\rm TF}^2/q^2$ in the limit of $q \to 0$ with $v_{\rm F}q/\omega \gg 1$ (the 
$q$-limit), where $\omega_{pl} (=\sqrt{4\pi e^2 n/m})$ is the plasmon energy at 
$q \to 0$ and $q_{\rm TF} (=\sqrt{4e^2 m p_{\rm F}/\pi})$ is the Thomas-Fermi 
screening constant. Associated with this singular point at the origin of 
$(q,\omega)$ space, the plasmon emerges as a collective mode satisfying 
$\varepsilon^R({\bm q},\omega) \approx 0$ at $\omega \approx \omega_{pl}$ for small $q$. 

In the dielectric-catastrophe regime ($r_s>r_s^c$), there appears the ghost exciton 
mode, but as long as $\tilde{\omega}_{\rm ex}({\bm q})$ is positive, the mode is 
not a pole on the real axis in the complex $\omega$-plane, indicating its irrelevance 
from a physical point of view. At both $q=0$ and $q_{\rm ex} (\neq 0)$ at which 
$\tilde{\omega}_{\rm ex}({\bm q})$ becomes zero [see Fig.~\ref{fig:11}(a)], however, 
the pole is situated on the real axis and thus it must have direct physical relevance. 

In fact, we find that $\varepsilon^R({\bm q},\omega)$ contains a couple of 
singularities, one at the origin of $(q,\omega)$ space and the other at $q=q_{\rm ex}$ 
and $\omega=0$, as illustrated in Fig.~\ref{fig:15}(b) for the case of $r_s=8$. The 
singularity occurring at the origin behaves in much the same way as that in the 
conventional metals, excluding a possibility of the observation of any exotic effects 
associated with this singularity. The only difference in the case of negative $\kappa$ 
from that of the conventional metals is seen in the sign of ${\rm Re}\, 
\varepsilon^R({\bm q},\omega)$ in the $q$-limit and concomitantly the reduction of 
the oscillator strength for the plasmon pole, but this reduction never becomes 
perfect (see Fig.~\ref{fig:12}), allowing the existence of the plasmon at any 
value of $r_s$. 

\begin{figure}[htbp]
\begin{center}
\includegraphics[scale=0.46,keepaspectratio]{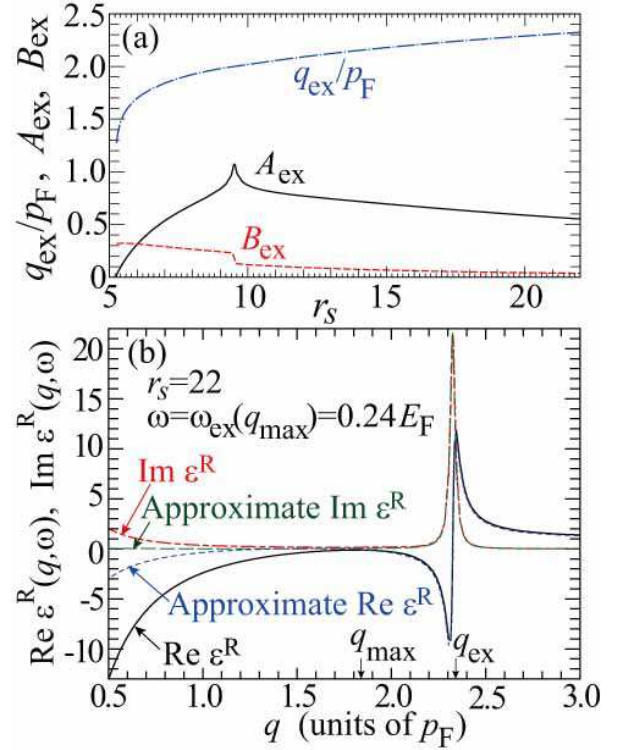}
\end{center}
\caption[Fig.16]{(Color online) (a) Calculated $q_{\rm ex}$ in units of $p_{\rm F}$, 
$A_{\rm ex}$, and $B_{\rm ex}$ as a function of $r_s$. (b) Comparison of 
$\varepsilon^R(q,\omega)$ between the full calculation in Eq.~(\ref{eq:53}) and the 
approximate one in Eq.~(\ref{eq:71}) at $r_s=22$ for $\omega=0.24E_{\rm F}$ at which 
energy $S({\bm q},\omega)$ becomes maximum due to the excitonic collective mode.}
\label{fig:16}
\end{figure}

The singularity at $q=q_{\rm ex}$ and $\omega=0$, on the other hand, has not been 
discussed in the literature and thus a detailed analysis is needed here. In the 
vicinity of this singularity, $\varepsilon^R({\bm q},\omega)$ is expressed as 
\begin{align}
\varepsilon^R({\bm q},\omega)\! \approx \!1\!+\!\left(\frac{p_{\rm F}}{q}\right )^2
\frac{1}{A_{\rm ex}(q\!-\!q_{\rm ex})/p_{\rm F}\!-\!iB_{\rm ex}\omega/E_{\rm F}},
\label{eq:71}
\end{align}
where the coefficients, $A_{\rm ex}$ and $B_{\rm ex}$, are given by 
$A_{\rm ex}=p_{\rm F}\partial f(q_{\rm ex},0)/\partial q_{\rm ex}$ and $B_{\rm ex}=
-\lim_{\omega \to 0}E_{\rm F}{\rm Im}f(q_{\rm ex},\omega)/\omega$ with $f(q,\omega)$, 
defined in reference to Eq.~(\ref{eq:53}) by 
\begin{align}
f(q,\omega)=\frac{\pi}{2}\frac{1}{\alpha r_s}\frac{1}{P_{\rm WI}(q,\omega)}
-\left (\frac{p_{\rm F}}{q}\right )^2G_s^R(q,\omega).
\label{eq:72}
\end{align}
Note that $f(q_{\rm ex},0)=0$ by the definition of $q_{\rm ex}$.

In Fig.~\ref{fig:16}(a), the obtained $q_{\rm ex}$, $A_{\rm ex}$, and $B_{\rm ex}$ are 
plotted as a function of $r_s$. At $r_s=r_s^c$, $A_{\rm ex}$ vanishes, removing this 
singularity from $\varepsilon^R({\bm q},\omega)$, but $A_{\rm ex}$ is positive definite 
for $r_s>r_s^c$ and thus this singularity always exists in the dielectric-catastrophe 
regime. With the increase of $r_s$, $q_{\rm ex}$ increases monotonically and at $r_s 
\approx 9.51$ it becomes equal to $2p_{\rm F}$, which is the boundary at the 
electron-hole single-pair excitation. Due to this transition across the boundary, 
both $A_{\rm ex}$ and $B_{\rm ex}$ exhibit anomalous behavior at $r_s \approx 9.51$. 
Because $B_{\rm ex}$ is much decreased at $r_s \approx 9.51$ and becomes smaller 
further for $r_s > 9.51$, the effect of this singularity on 
$\varepsilon^R({\bm q},\omega)$ is gradually enhanced with the increase of $r_s$. 

In Fig.~\ref{fig:16}(b), we have given an example of the comparison between the 
approximate form for $\varepsilon^R({\bm q},\omega)$ in Eq.~(\ref{eq:71}) and 
the full result of $\varepsilon^R({\bm q},\omega)$ in Eq.~(\ref{eq:53}) calculated 
at $r_s=22$, from which we can confirm the accuracy of this approximate form 
in the neighborhood of the singular point at $(q,\omega)\!=\!(q_{\rm ex},0)$. 
This figure also shows that the excitonic-peak position providing the maximum height 
in $S({\bm q},\omega)$, $(q_{\rm max},\omega_{\rm ex}(q_{\rm max}))$, is actually 
coincident with the point in $(q,\omega)$ space at which 
$\varepsilon^R({\bm q},\omega) \approx 0$, validating to call $\omega_{\rm ex}(q)$ 
``the excitonic collective mode''. In this way, the excitonic collective mode is 
brought about by the singular behavior of $\varepsilon^R({\bm q},\omega)$ 
associated with the singularity at $(q_{\rm ex},0)$ which is originally induced 
by the ghost exciton mode. 

\begin{figure}[thbp]
\begin{center}
\includegraphics[scale=0.46,keepaspectratio]{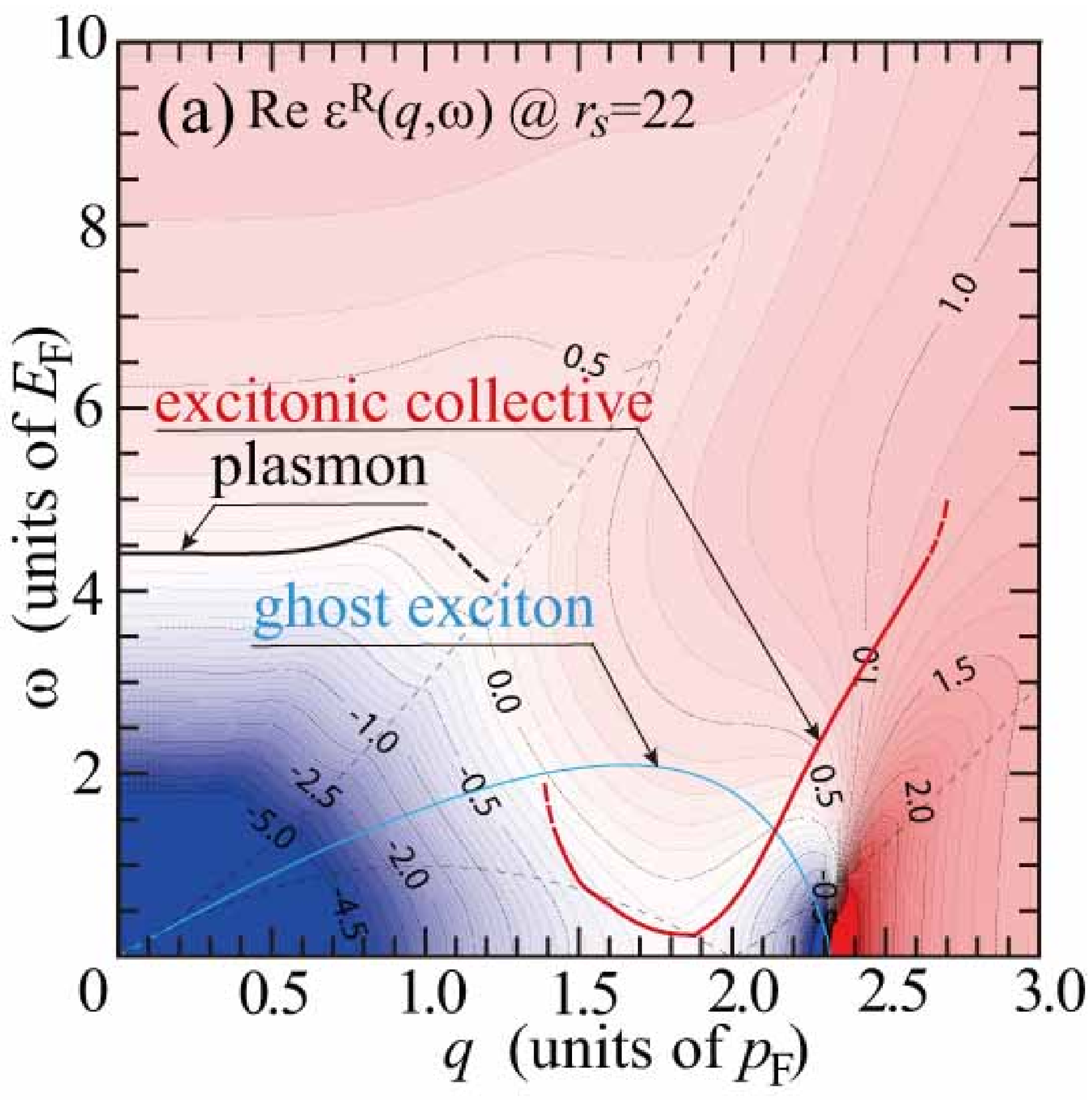}
\includegraphics[scale=0.46,keepaspectratio]{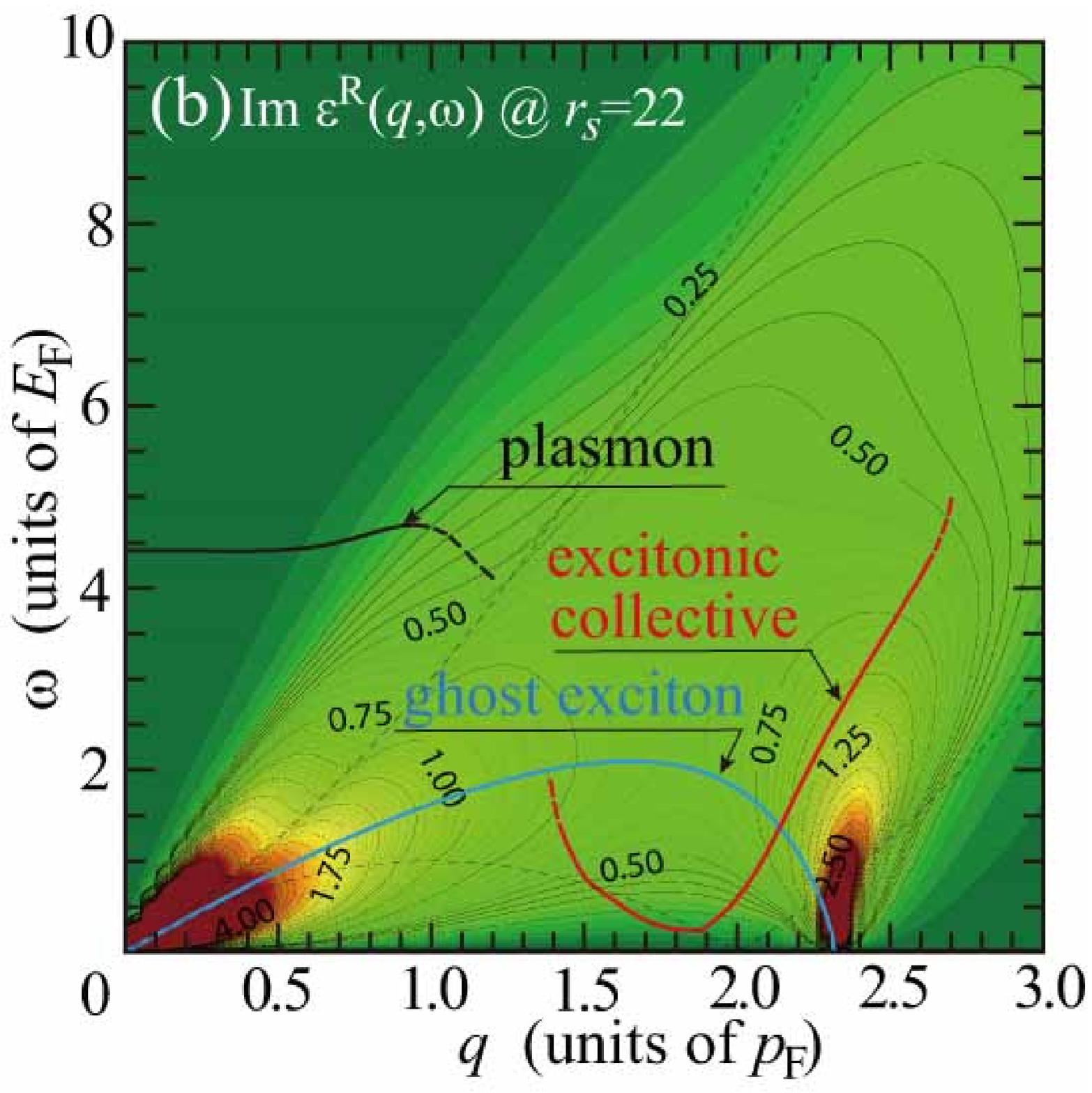}
\end{center}
\caption[Fig.17]{(Color online) Two-dimensional contour map of 
$\varepsilon^R({\bm q},\omega)$ at $r_s=22$ for (a) real and (b) imaginary parts.}
\label{fig:17}
\end{figure}

Finally in Fig.~\ref{fig:17}, $\varepsilon^R({\bm q},\omega)$ given in 
Eq.~(\ref{eq:53}) is plotted at $r_s=22$ in the whole $(q,\omega)$ space in order 
to fully represent the typical behavior of $\varepsilon^R({\bm q},\omega)$ in the 
dielectric-catastrophe regime. As a general feature, $\varepsilon^R({\bm q},\omega)$ 
contains a couple of singular points. Associated with each singularity, there 
appears a collective mode at which $\varepsilon^R({\bm q},\omega) \approx 0$, 
leading to the double-peak structure in $S({\bm q},\omega)$, one for the plasmon 
and the other for the excitonic collective mode, as shown in Fig.~\ref{fig:1}(b). 

\section{Conclusion and Discussion}
\label{sec:4}

In this paper, we have calculated the momentum distribution function $n({\bm p})$ 
and the dynamical structure factor $S({\bm q},\omega)$ in the three-dimensional 
(3D) electron gas at low densities, i.e., $r_s>5.25$, in which the electronic 
compressibility $\kappa$ is negative. The obtained $n({\bm p})$ is considered to be 
sufficiently accurate, because it rigorously satisfies not only the exact 
asymptotic behavior but also the three sum rules, Eqs.~(\ref{eq:3})-(\ref{eq:5}), 
the last of which is successfully derived in this work. The obtained 
$S({\bm q},\omega)$ is also considered to be sufficiently accurate, because the two 
ingredients to construct $S({\bm q},\omega)$, namely, $\Pi_{\rm WI}^R({\bm q},\omega)$ 
and $G_s^R({\bm q},\omega)$, are both reliably determined in accordance with all the 
known constraints that must be fulfilled for correct evaluation of those quantities. 
For $r_s>10$, the calculated $S({\bm q},\omega)$ exhibits a peak structure due to 
the excitonic collective mode in addition to the well-known plasmon peak. With the 
increase of $r_s$, this new peak structure grows steadily and eventually for 
$r_s>20$ it dominates the plasmon peak. Associated with this excitonic collective 
mode, a singularity is found for the first time in $\varepsilon^R({\bm q},\omega)$ 
at $q \approx 2p_{\rm F}$ and $\omega=0$, which is the direct physical consequence of 
the appearance of the ghost exciton mode for $r_s>5.25$. 

Four comments are in order: 

(i) Although we have treated the 3D electron gas in this paper, exactly the same 
physics will be found in the 2D electron gas in which the dielectric catastrophe 
regime appears for $r_s\!>\!2$. Actually, negative $\kappa$ has already been 
observed in the 2D system.~\cite{Eisenstein1992,Ilani2000} Thus, 
in order to confirm the emergence of the excitonic collective mode by the 
measurement of $S({\bm q},\omega)$ in some suitably designed experiments, the 2D 
system may be more recommended, although detailed calculation for the 2D system is 
left for the future. 

(ii) Physically, if the excitonic collective mode is a dominant polarization process 
in the charge response to an external point-charge perturbation, we may imagine that 
the screening effect will be much reduced than that in the usual metals due to the 
fact that the excitation of tightly-bound electron-hole pairs does not contribute to 
screening, indicating that the Hartree-Fock (HF) approximation may work well in the 
evaluation of some of physical quantities. This might be the reason why Ilani et 
al.~\cite{Ilani2000} found that HF described their experimental results well in the 
metallic state at negative $\kappa$. 

(iii) As mentioned in Secs.~\ref{sec:2F} and \ref{sec:3E}, we find indications of 
some exotic electronic phase transition of the 3D electron gas for $r_s\!>\!20$. We 
need to make a further study of the possibilities of such a phase transition, but at 
the same time, we need to take account of a possibility of partially spin-polarized 
state for $20\!<\!r_s\!<\!40$,~\cite{Ortiz1999} suggesting us to investigate the 
fate of the excitonic collective mode in the spin-polarized state as well as the 
coupling between charge and spin channels, a difficult but very intriguing problem 
left in the future. 

(iv) Finally, we remark on similarity and difference between the plasma and the 
excitonic collective modes. Due to $\varepsilon^R({\bm q},\omega)\! \approx \! 0$ at 
each mode in the relation of ${\bm D}({\bm q},\omega)=\varepsilon^R({\bm q},\omega)
{\bm E}({\bm q},\omega)$, there exists a wave-like charge-density oscillation as an 
eigenmode, accompanied by a finite electric field ${\bm E}({\bm q},\omega)$ 
even in the absence of an external force (i.e., ${\bm D}({\bm q},\omega)\!=
\!{\bm E}({\bm q},\omega)\!+\!4\pi{\bm P}({\bm q},\omega)\!\approx\!0$). The 
associated electric polarization ${\bm P}({\bm q},\omega)$ is made of collective 
electron-hole excitations and works as a restoring force. Those charge-density 
oscillations can be quantized as bosons, just as the plasma oscillations quantized 
into the plasmons. In spite of those similarities, there is an important difference 
in the correlation among excited electrons and holes; for the plasmons, because 
its energy is larger than $E_{\rm F}$, all excited electrons and holes are 
uncorrelated, as can be well treated in RPA. For the excitonic collective modes, 
on the other hand, the excitations occur as a collection of excitons, indicating 
no polarization field in the long-wavelength limit and thus the absence of the 
mode at $|{\bm q}|\!\to\! 0$; the polarization appears, if $|{\bm q}|$ is of the 
order of $2p_{\rm F}$ or the inverse of the exciton binding radius. Incidentally, 
if we consider the exotic phase mentioned in (iii) in terms of a spontaneous 
excitation of this charge-density oscillation, it might be regarded as a CDW state 
in the sense of not nesting-driven but Overhauser~\cite{Overhauser,Perdew}.


\acknowledgments

This work is supported by a Grant-in-Aid on Innovative Area "Materials Design 
through Computics: Complex Correlation and Non-Equilibrium Dynamics" (No. 22104011) 
from the Ministry of Education, Culture, Sports, Science, and Technology, Japan. 


\end{document}